\documentclass[3p,times,twocolumn]{elsarticle}
\usepackage{ecrc}
\volume{00}
\firstpage{1}
\journalname{Nuclear Physics B Proceedings Supplement}
\runauth{}
\jid{nuphbp}
\jnltitlelogo{Nuclear Physics B Proceedings Supplement}
\usepackage{amssymb}
\biboptions{sort&compress}

\pdfmapfile{+txfonts.map}
\usepackage{epstopdf}
\def\be{\begin{equation}}
\def\ee{\end{equation}}
\def\beqn{\begin{eqnarray}}
\def\eeqn{\end{eqnarray}}
\def\no{\nonumber}
\def\ba{\begin{array}{c}}
\def\bat{\begin{array}{cc}}
\def\ea{\end{array}}
\def\bi{\begin{itemize}}
\def\ei{\end{itemize}}

\def\cL{{\cal L}}

\def\cO{{\cal O}}

\def\cY{{\cal Y}}

\newcommand{\eqn}[1]{(\ref{#1})}
\newcommand{\bel}[1]{\be\label{#1}}
\begin{document}

\begin{frontmatter}

%% Title, authors and addresses

%% use the tnoteref command within \title for footnotes;
%% use the tnotetext command for the associated footnote;
%% use the fnref command within \author or \address for footnotes;
%% use the fntext command for the associated footnote;
%% use the corref command within \author for corresponding author footnotes;
%% use the cortext command for the associated footnote;
%% use the ead command for the email address,
%% and the form \ead[url] for the home page:
%%
%% \title{Title\tnoteref{label1}}
%% \tnotetext[label1]{}
%% \author{Name\corref{cor1}\fnref{label2}}
%% \ead{email address}
%% \ead[url]{home page}
%% \fntext[label2]{}
%% \cortext[cor1]{}
%% \address{Address\fnref{label3}}
%% \fntext[label3]{}

\dochead{}
%% Use \dochead if there is an article header, e.g. \dochead{Short communication}

\title{ICHEP 2014 Summary: Theory Status after the First LHC Run}
%   ICHEP 2014: Theory Summary and Outlook}

%% use optional labels to link authors explicitly to addresses:
%% \author[label1,label2]{<author name>}
%% \address[label1]{<address>}
%% \address[label2]{<address>}

\author{Antonio Pich}

\address{IFIC, Universitat de Val\`encia -- CSIC, Apt. Correus 22085, E-46071 Val\`encia, Spain}

\begin{abstract}
A brief overview of the main highlights discussed at ICHEP 2014 is presented. The experimental data confirm that the scalar boson discovered at the LHC couples to other particles as predicted in the Standard Model. This constitutes a great success of the present theoretical paradigm, which has been confirmed as the correct description at the electroweak scale. At the same time, the negative searches for signals of new phenomena
tightly constrain many new-physics scenarios, challenging previous theoretical wisdom and opening new perspectives in fundamental physics. Fresh ideas are needed to face the many pending questions unanswered within the Standard Model framework.
\end{abstract}

\begin{keyword}
%% keywords here, in the form: keyword \sep keyword

%% MSC codes here, in the form: \MSC code \sep code
%% or \MSC[2008] code \sep code (2000 is the default)
High Energy Physics \sep Electroweak and Strong Interactions \sep Standard Model and Beyond
\end{keyword}

\end{frontmatter}

%%
%% Start line numbering here if you want
%%
% \linenumbers

%% main text
\section{Introduction}
\label{sec:Introduction}

The large number of results discussed at this conference
exhibit
% can be simply summarized as
an overwhelming success of the Standard Model (SM).
Combined with all previous precision tests, the first LHC run has finally established the whole SM framework as the true theory of the electroweak interactions at the experimentally accessible energy scales. The new boson discovered at 125 GeV \cite{Aad:2012tfa,Chatrchyan:2012ufa} behaves indeed as the expected SM scalar and its mass fixes the last free parameter of the Lagrangian.

At the same time, all LHC searches for exotic objects have given negative results, pushing the scale of new physics well above the TeV and putting in trouble the most fashionable theoretical scenarios for physics beyond the SM. Many models are either already ruled out or its surviving parameter space is no-longer appropriate to address the physics problems they were supposed to solve.
Clearly, the LHC constraints should imply major changes in our theoretical guidelines/prejudices when searching for new-physics explanations to the many open questions that the SM leaves unanswered.

The following sections briefly summarize the present status, emphasizing those aspects
more closely related to the recent experimental developments.

\section{Precision QCD}
\label{sec:QCD}

The success of the experimental LHC program relies to a large extent in the existence of precise theoretical predictions for the relevant signals and the many QCD backgrounds. The last few years have witnessed an impressive progress in the theoretical accuracy, with many perturbative calculations reaching already the next-to-next-to-leading order (NNLO) plus re-summation of leading logarithms (LL) at the NNLL \cite{Nason,Uwer}. The on-going effort towards N${}^k$LO accuracy \cite{Anastasiou:2015ema}
is complemented by a corresponding improvement of the parton distribution functions, updated Monte Carlo generators with appropriate matching of matrix elements and parton showering, and more efficient tools to address multi-particle interactions at higher orders. The measurement of many different cross sections \cite{Roda,Carli,Berryhill}, spanning a broad range from $10^{-3}$ to $10^{6}$ pb and in remarkable agreement with their predicted SM values, provides a very significant validation of the adopted theoretical framework.

QCD has been beautifully tested, verifying at the four-loop level the running of its unique coupling over all accesible energies, from the $\tau$ mass \cite{Pich:2013lsa} to the highest momentum scales reached at colliders \cite{Malaescu:2012ts,Chatrchyan:2013txa}.
Although several new determinations of the strong coupling have been recently included in the world average, its final value remains
% stable: $\alpha_s(M_Z^2)=0.1185\, (6)$ \cite{Agashe:2014kda,Pich:2013sqa}.
stable \cite{Agashe:2014kda,Pich:2013sqa}:
\bel{eq:alpha}
\alpha_s(M_Z^2)\, =\, 0.1185\pm 0.0006\, .
\ee
The lattice result, which has the smallest assigned uncertainty, is now reinforced with contributions from several groups. The lattice average agrees with the average of all other results, but it largely determines the final error. Recent lattice simulations provide also more precise values of the $u, d, s, c$ and $b$ quark masses \cite{ElKhadra,Aoki:2013ldr}.

At low energies, the confining nature of the QCD forces makes more difficult to perform accurate predictions. Nevertheless, the combined use of effective field theory tools, largely based on symmetry considerations, and lattice simulations in powerful computer mainframes has made possible to achieve an impressive progress in recent years \cite{Aoki:2013ldr}. This is, however, not yet enough to quantitatively understand the wealth of new exotic states which have been experimentally identified in the heavy-quark spectroscopy \cite{Brambilla:2014jmp,Peng}.

The LHC experiments are collecting a rich harvest of heavy-ion collisions, complementing at higher energies the successful RHIC studies. Beautiful signals of jet quenching, screening ($J/\Psi,\,\Upsilon$) and regeneration ($J/\Psi$) have been obtained \cite{Wessels}. Moreover, there is now clear evidence for collective phenomena in proton-lead collisions also. A standard picture of high-energy heavy-ion collisions is emerging where relativistic hydrodynamics seems to play a prominent role, the quark-gluon plasma behaving as a near-perfect relativistic fluid \cite{Gale}.

Promising developments are also arising from the more formal investigations of conformal field theories (CFT), strings, dualities, etc. \cite{Barbon}. The large symmetries present in supersymmetric Yang-Mills theories make easier to obtain results, from which one hopes to learn something about the much more difficult QCD case. AdS/CFT ideas are already being applied to heavy-ion and condense-matter physics. Moreover, they are changing the way one looks at gravity with the emergence of space out of quantum mechanics through the holography concept. Last but not least, intricate mathematics are being uncovered (twistor diagrams, on-shell recursion relations, permutations and Yangian symmetry, etc.), which should help to understand the astonishing simplicity of many results, obtained through complicated perturbative calculations with Feynman diagrams, pointing the way to more efficient algorithms.

\section{The Heaviest Mass Scale}
\label{sec:Top}

The top quark is a very sensitive probe of the electroweak symmetry breaking (EWSB), since it is the heaviest fundamental particle within the SM framework. Its Yukawa coupling is very close to one,
\bel{eq:TopYukawa}
y_t\, = \,\frac{\sqrt{2}}{v}\, m_t\, =\, 2^{3/4} G_F^{1/2} m_t \,\approx\, 1\, .
\ee
Therefore, virtual top contributions dominate the electroweak quantum corrections in many relevant observables. The large value of $m_t$ makes the top very different from all other quarks ($y_b\approx 0.025$, $y_c\approx 0.007 \gg y_{s,d,u}$).
One could then wonder whether it is really a genuine SM particle. If some (non-perturbative) strong dynamics is responsible for the EWSB, the top should obviously be directly linked to it.

%So far, we have only detected top quarks through their decay mode $t\to W^+ b$, because
%the top couplings to the lighter quark generations are very small. From the measured single-top production cross section, one extracts
%$|V_{tb}| > 0.92 \; (95\%\,\mathrm{CL})$ \cite{Aaltonen:2015cra,Khachatryan:2014iya}.
So far, we have only detected top quarks through their decay mode $t\to W^+ b$, because
the top couplings to the lighter quark generations are very small. The measured single-top production cross section implies \cite{Aaltonen:2015cra,Khachatryan:2014iya}:
\bel{eq:Vtb}
|V_{tb}|\, >\, 0.92 \qquad (95\%\,\mathrm{CL}) .
\ee

The existence of a small discrepancy between the measured
forward-backward asymmetry in $t\bar t$ production at the Tevatron and the SM prediction \cite{Kuhn:2011ri} has triggered many suggestions for new-physics explanations.
Additional higher-order corrections are found to be small and there is good agreement in the theoretical SM result. However, the most recent measurements, based on the full data samples, significantly lower the discrepancy, particularly at D0 \cite{Abazov:2014cca,Bernardi} where they are now consistent with the SM. In $pp$ collisions one measures a corresponding charge asymmetry, arising from the annihilation of valence quarks with sea antiquarks. The LHC data is in good agreement with the SM.

The collider measurements of the top mass are getting quite accurate \cite{Carli,Bernardi,Aldaya}. The last world combination has a precision of $0.44\%$
\cite{ATLAS:2014wva}:
\bel{eq:TopMass}
m_t^{\mathrm{MC}} = (173.34\pm 0.76)\;\mathrm{GeV}\, .
\ee
However, this value is obtained from a kinematical reconstruction of the top decay products and refers to the mass parameter implemented in the Monte Carlo simulations.
Its relation with a well-defined QCD mass in a given renormalization scheme is still unclear. In most precision electroweak tests where $m_t$ is needed as input, the result \eqn{eq:TopMass} is identified with the so-called pole mass, the pole of the perturbative quark propagator, which would correspond to the on-shell mass of a free fermion. Confinement implies that this is a badly defined quantity and the uncertainty associated with this naive assumption is controversial. The most recent studies estimate an additional theoretical uncertainty from this source of the order of 1 GeV \cite{Hoang:2008xm,Moch:2014tta,Uwer}.

A well-defined top mass can be obtained from the $t\bar t$ production cross section, which has been theoretically computed to NNLO + NNLL accuracy \cite{Uwer}. The present Tevatron and LHC determinations are consistent with Eq.~\eqn{eq:TopMass} but have much larger uncertainties. % of about 3 GeV.
The most precise result so far has been obtained by ATLAS \cite{Aad:2014kva}:
\bel{eq:TopMass2}
m_t^{\mathrm{pole}} = (172.9\, {}^{+\, 2.5}_{-\, 2.6})\;\mathrm{GeV}\, .
\ee
Different methods to improve the sensitivity are being suggested \cite{Moch:2014tta,Alioli:2013mxa,ATLAS-mt,Kawabataa:2014osa}, and a precision of 1~GeV seems possible at the LHC. A further order of magnitude improvement would require to measure  $\sigma(e^+e^-\to t\bar t)$ in the threshold region, at future $e^+e^-$ colliders.

\section{A New Higgs-like Boson}
\label{sec:Higgs}

The new boson discovered at the LHC complies with the expected behaviour, with a spin/parity consistent with the SM $0^+$ assignment \cite{Bernardi,Kado,David}. The observation of its $2\gamma$ decay mode demonstrates that it is a boson with $J\not= 1$, while the $J^P = 0^-$ and $2^\pm$ hypotheses are already excluded at confidence levels above 99\%. The masses measured by ATLAS and CMS are in good agreement, giving the average value \cite{Aad:2015zhl}
\bel{eq:M_H}
M_H = (125.09\pm 0.24)\;\mathrm{GeV}\, .
\ee

An important question to address is whether $H$ corresponds to the unique scalar boson incorporated in the SM, or it is just the first signal of a much richer scenario of EWSB \cite{Grojean}. Obvious possibilities are an extended scalar sector with additional fields or dynamical (non-perturbative) EWSB generated by some new underlying dynamics. Whatever the answer turns out to be, the LHC finding represents a truly fundamental discovery with far reaching implications. If it is an elementary scalar (the first one), one would have established the existence of a bosonic field (interaction) which is not a gauge force. If instead, it is a composite object,
a completely new underlying interaction should exist.

A fundamental scalar requires some protection mechanism to stabilize its mass. If there is new physics at some heavy scale $\Lambda_{\mathrm{NP}}$, quantum corrections could bring the scalar mass $M_H$ to the new-physics scale $\Lambda_{\mathrm{NP}}$:
\bel{eq:MH-corrections}
\delta M_H^2\,\sim\, \frac{g^2}{(4\pi)^2}\, \Lambda_{\mathrm{NP}}^2\,\log{(\Lambda_{\mathrm{NP}}^2/M_H^2)}\, .
\ee
Which symmetry keeps $M_H$ away from $\Lambda_{\mathrm{NP}}$? Fermion masses are protected by chiral symmetry, while gauge symmetry protects the gauge boson masses; those particles are massless when the symmetry becomes exact.
Supersymmetry was originally advocated to protect the Higgs mass, but according to present data this no-longer works `naturally'. Another possibility would be scale symmetry, which in the SM is broken by the Higgs mass; a naive dilaton is basically ruled out, but there could be an underlying conformal theory at $\Lambda_{\mathrm{NP}}$. Dynamical EWSB with light pseudo-Goldstone particles at low energies remains also a viable scenario.
Future discoveries at the LHC should bring a better understanding of the correct dynamics above the electroweak scale.

\subsection{SM EWSB Mechanism}
\label{subsec:SM}

The SM incorporates a complex scalar doublet $\Phi$ with a non-trivial potential which triggers the wanted EWSB:
\begin{equation}\label{eq:Lphi}
\cL_\Phi\; =\; (D_\mu\Phi)^\dagger D^\mu\Phi - \lambda\, \left( |\Phi|^2 -\frac{v^2}{2}\right)^2 \, .
\end{equation}
In the unitary gauge the three Goldstone fields give rise to the longitudinal polarizations of the $W^\pm$ and $Z$, generating their masses through the derivative term in (\ref{eq:Lphi}): $M_W = M_Z \cos{\theta_W} = g v/2$. A massive scalar field $H$, the Higgs boson, remains because $\Phi$ contains a fourth degree of freedom, which is not needed for the EWSB
\cite{Higgs:1964pj,Englert:1964et,Guralnik:1964eu,Kibble:1967sv}. The scalar doublet structure provides a renormalizable model with good unitarity properties.

%%%%%%%%%%%%%%%%%%%%%% Figure EW fit %%%%%%%%%%%%%%%%%%%%%
\begin{figure}[t]
%\centering
%\mbox{}\hskip -.05cm
\includegraphics[width=\columnwidth,clip]{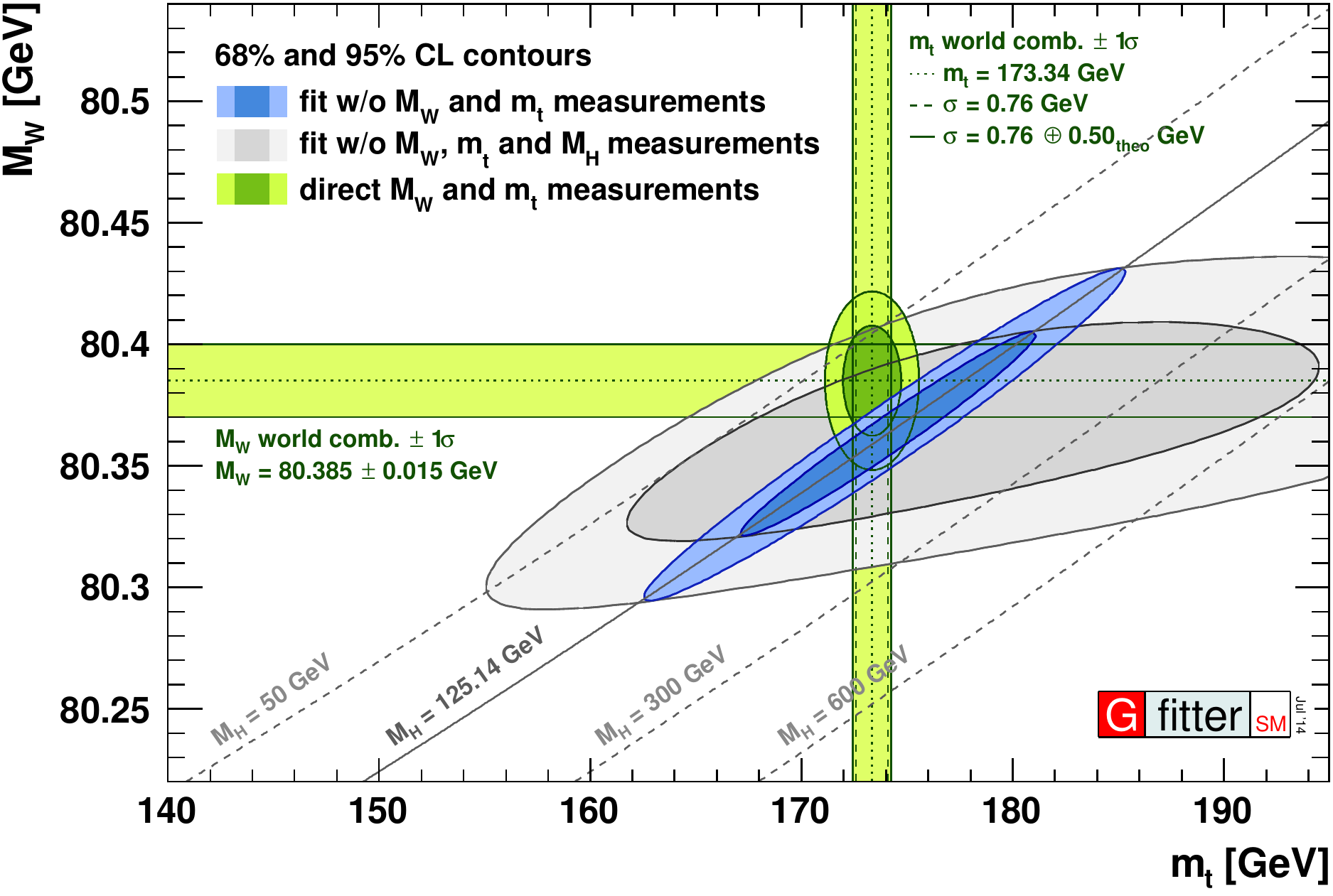} % 7.5cm
\caption{SM electroweak fit in the $m_t$--$M_W$ plane, with (blue) and without (gray) the Higgs mass, compared with the direct measurements of the top and $W$ masses (green) \cite{Baak:2014ora}.}
\label{fig:EWfit_mt}
\end{figure}
%%%%%%%%%%%%%%%%%%%%%%%%%%%%%%%%%%%%%%%%%%%%%%%%%%%%%%%%%%%%%%%%%%%

While the vacuum expectation value (the electroweak scale) was already known,
$v = (\sqrt{2} G_F)^{-1/2} = 246$~GeV, the measured Higgs mass in Eq.~\eqn{eq:M_H} determines the last free parameter of the SM, the quartic scalar coupling:
\begin{equation}\label{eq:lambda}
\lambda\; =\; \frac{M_H^2}{2 v^2}\; =\; 0.13\, .
\end{equation}
The experimental value of $M_H$ is in beautiful agreement with the expectations from global fits to precision electroweak data \cite{Baak:2014ora,Freitas},
shown in Fig.~\ref{fig:EWfit_mt}.
Notice that the green vertical band in the figure assumes that $m_t^{\mathrm{MC}}$ in Eq.~\eqn{eq:TopMass} is the pole top mass.

\begin{figure}[t]
\centering
\includegraphics[width=7.55cm,clip]{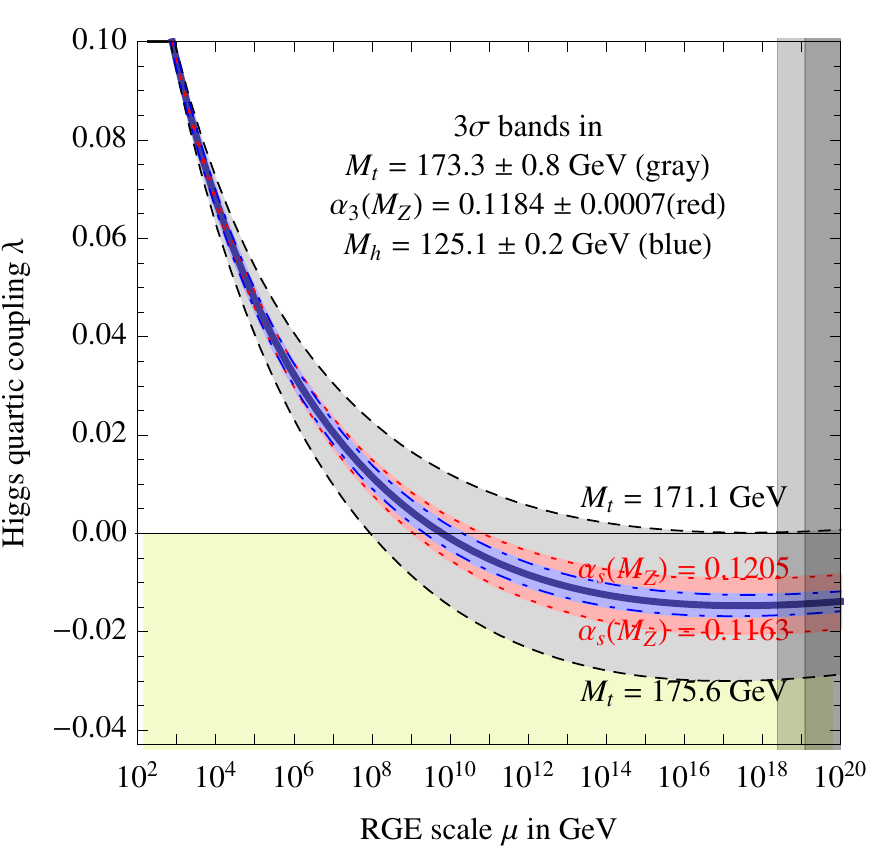} %7.15
\caption{Evolution of $\lambda(\mu)$ with the renormalization scale  \cite{Buttazzo:2013uya}.}
\label{fig:lambda}
\end{figure}

Quantum corrections to $M_H^2$ are dominated by contributions from heavy top loops, which grow logarithmically with the renormalization scale $\mu$:
\bel{eq:MH_QC}
\frac{M_H^2}{2 v^2} \approx \lambda(\mu) + \frac{2 y_t^2}{(4\pi)^2}\left[\lambda + 3 (y_t^2-\lambda)\,\log{(\mu/m_t)}\right] .\;
\ee
As expected, $M_H$ is brought close to the heaviest SM mass $m_t= y_t v/\sqrt{2}$.
Since the physical value of
$M_H$ is fixed, the tree-level contribution $2 v^2\lambda(\mu)$ decreases with increasing $\mu$. Fig.~\ref{fig:lambda} shows the evolution of $\lambda(\mu)$ up to the Planck scale ($M_{\mathrm{Pl}} = 1.2\times 10^{19}$~GeV), at the NNLO, varying $m_t$, $\alpha_s(M_Z^2)$ and $M_H$ by $\pm 3\sigma$ \cite{Buttazzo:2013uya}. The quartic coupling remains weak in the entire energy domain below $M_{\mathrm{Pl}}$ and crosses $\lambda=0$ at very high energies, around $10^{10}$~GeV. The values of $M_H$ and $m_t$ are very close to those needed for absolute stability of the potential ($\lambda >0$) up to  $M_{\mathrm{Pl}}$, which would require $M_H > (129.6\pm 1.5)$~GeV \cite{Buttazzo:2013uya}
($\pm 5.6$~GeV with more conservative errors on $m_t$ \cite{Alekhin:2012py}). Even if $\lambda$ becomes slightly negative at very high energies, the resulting potential instability leads to an electroweak vacuum lifetime much larger than any relevant astrophysical or cosmological scale. Thus, the measured Higgs and top masses result in a metastable vacuum \cite{Buttazzo:2013uya} and the SM could be valid up to $M_{\mathrm{Pl}}$. The possibility of some new-physics threshold at scales $\Lambda\sim M_{\mathrm{Pl}}$, leading to the matching condition
 $\lambda(\Lambda) = 0$, is obviously intriguing.

%%%%%%%%%%%%%%%%%%%%%%% Figure EW fit %%%%%%%%%%%%%%%%%%%%%
%%%%%%%%%%%%%%%%%%%%%%%%%%%%%%%%%%%%%%%%%%%%%%%%%%%%%%%%%%%%%%%%%%%%

\subsection{Higgs Couplings}

The new scalar appears to couple to the fermions and gauge bosons ($W^\pm$, $Z$, $\gamma$, $G^a$) with the strength expected for the SM Higgs.
The sensitivity to the different couplings is increased disentangling the possible production channels: gluon fusion ($GG\!\to\! t\bar t\!\to\! H$), vector-boson fusion ($VV\!\to\! H$, $V\! \! =W,Z$) and associated $VH$ or $t\bar t H$ production. At the LHC gluon-fusion dominates, giving access to the top Yukawa.
The measured $H$ production cross section confirms the existence of a $t\bar tH$ coupling with the SM size. Moreover, it excludes additional fermionic contributions to gluon-fusion production; a fourth quark generation would increase the cross section by a factor of nine, and much larger enhancements would result from exotic fermions in higher colour representations, coupled to the Higgs~\cite{Ilisie:2012cc}.

%%%%%%%%%%%%%%%%%%%%%%% Figure Higgs couplings %%%%%%%%%%%%%%%%%%%%%
\begin{figure}[t]
\centering
\includegraphics[width=7.4cm,clip]{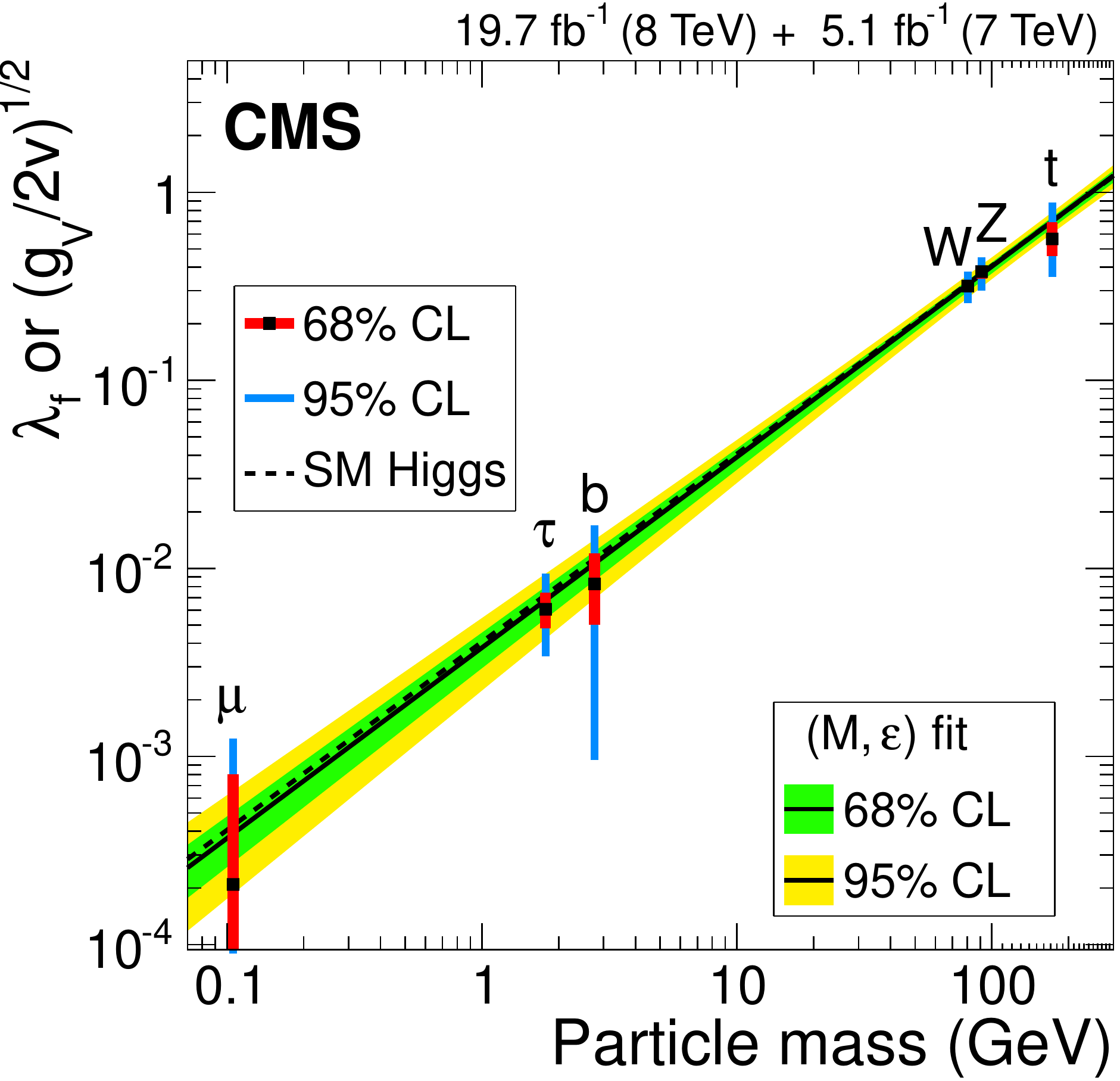} %7.0
\caption{Higgs couplings to different particles %as function of their masses
\cite{Khachatryan:2014jba}.}
\label{fig:Hcouplings}
\end{figure}
%%%%%%%%%%%%%%%%%%%%%%%%%%%%%%%%%%%%%%%%%%%%%%%%%%%%%%%%%%%%%%%%%%%%

The decay $H\to\gamma\gamma$ occurs in the SM through intermediate $W^+W^-$ and $t\bar t$ triangular loops, which interfere destructively.
Therefore, it is sensitive to new physics contributions such as charged scalar loops. The LHC data \cite{Khachatryan:2014jba,Aad:2014eha} favour the SM sign for the top Yukawa and strongly constrain any additional contributions beyond the SM.
The tree-level decays $H\to W^{\pm *} W^{\mp}, Z^* Z$
%V^*V$ ($V=W^\pm,Z$)
directly test the electroweak gauge couplings of the $H$ boson \cite{Khachatryan:2014jba,Aad:2014eva,ATLAS:2014aga}. In addition, we have now strong evidence for the $H$ coupling to $b\bar b$ and $\tau^+\tau^-$ through the corresponding fermionic decays \cite{Khachatryan:2014jba,Aad:2014xzb,Aad:2015vsa}.

The mass dependence of the $H$ couplings has been clearly verified by the data, as shown in Fig.~\ref{fig:Hcouplings}. Fitting the measured couplings with the parametrization \cite{Ellis:2013lra}
$y_f  = \sqrt{2}\, (m_f/M)^{1+\epsilon}$\ and\
$(g_{HVV}^{\phantom{2}}/2v)^{1/2} = (M_V/M)^{1+\epsilon}$, one gets
$M\in [217,\, 279]$~GeV and $\epsilon\in [-0.054,\, 0.100]$ (95\% CL) \cite{Khachatryan:2014jba}, in agreement with the SM values $(M,\,\epsilon) =(v,\, 0)$. Moreover, the 95\% CL upper limit $\mathrm{Br}(H\to e^+e^-) < 1.9\times 10^{-3}$ \cite{Khachatryan:2014jba} confirms the predicted suppression of the electronic coupling.

\section{Flavour Physics}
\label{sec:Flavour}

%%%%%%%%%%%%%%%%%%%%%%% CKM fit %%%%%%%%%%%%%%%%%%%%%%%%%%%%%%%%%%%%
\begin{figure}[t]
\centering
\includegraphics[width=\columnwidth,clip]{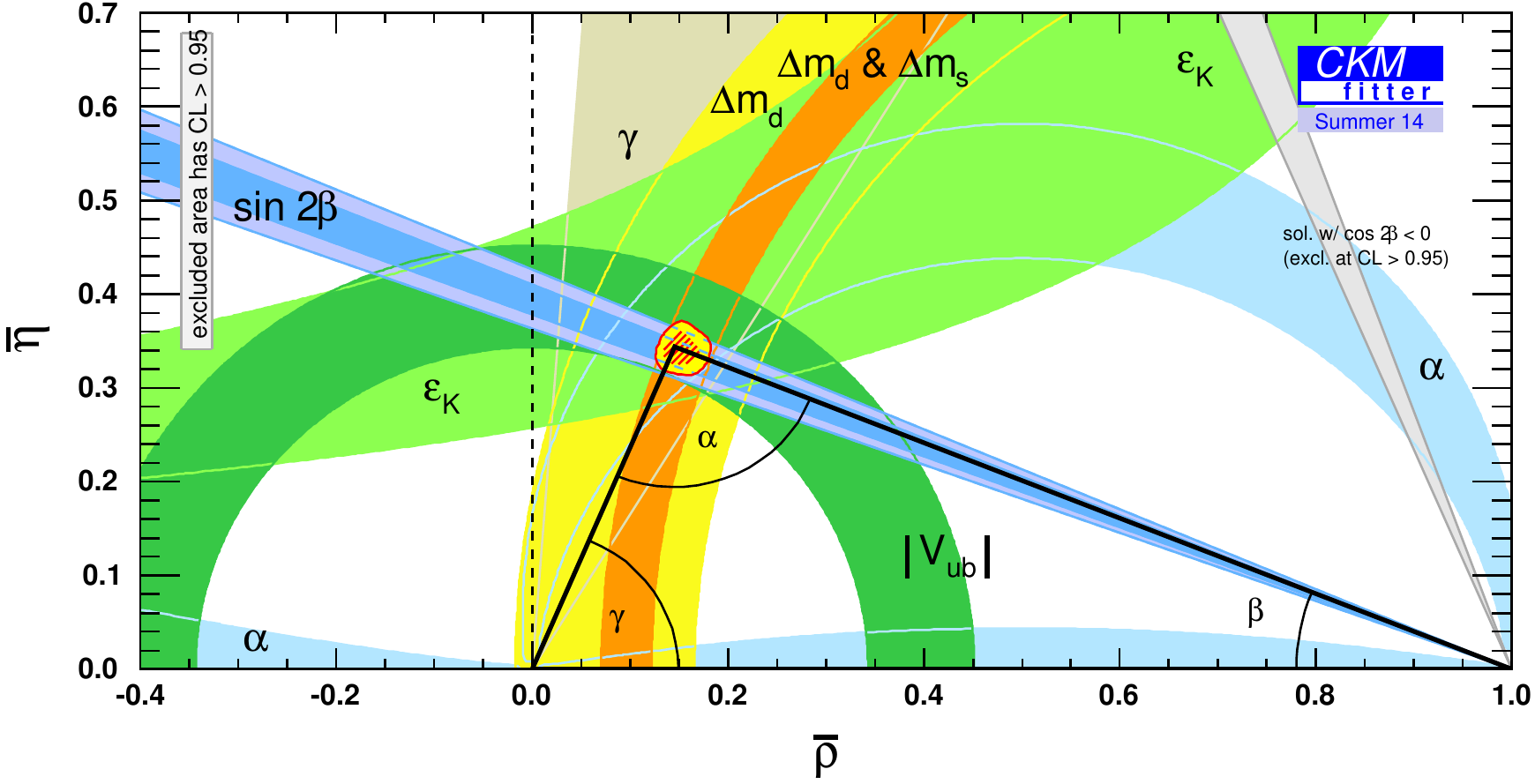}
\caption{Experimental constraints on the SM unitarity triangle \cite{Charles:2015gya}.}
\label{fig:CKMfit}
\end{figure}
%%%%%%%%%%%%%%%%%%%%%%%%%%%%%%%%%%%%%%%%%%%%%%%%%%%%%%%%%%%%%%%%%%%%

The fermion masses and mixings are all determined by the Yukawa couplings, which are matrices in flavour space. Their diagonalization leads to a three-generation quark mixing matrix $V_{u_i d_j}$, involving four measurable parameters: three angles and one CP-violating phase. While the SM does not provide any real understanding of flavour,
it accommodates all quark-flavour phenomena studied so far \cite{Browder,Albrecht} through this unitary mixing matrix with a characteristic hierarchical structure \cite{Wolfenstein:1983yz}:
\beqn
\hskip -.65cm
V &\!\!\! = & \!\!\!
{\setlength{\arraycolsep}{5pt}
\left[ \begin{array}{ccc}
1\! -\! {\lambda^2 \over 2} & \lambda & A\lambda^3 (\rho \! -\! i\eta)
\\
-\lambda & 1 \! -\! {\lambda^2 \over 2} & A\lambda^2
\\
A\lambda^3 (1\! -\!\rho \! -\! i\eta) & -A\lambda^ 2 &  1
\ea\right]}
\, +\, O\!\left(\lambda^4 \right) \, .
\nonumber\eeqn
A global fit to flavour data gives
$\lambda = 0.22548\,{}^{\raisebox{1pt}{$\scriptstyle +\, 0.00068$}}_{\raisebox{2pt}{$\scriptstyle -\, 0.00034$}}$,
% $\lambda = 0.22548\,{}^{+\, 0.00068}_{-\, 0.00034}$,
$A=0.810\,{}^{+\, 0.018}_{-\, 0.024}$, $\bar\rho = 0.145\, {}^{+\, 0.013}_{-\, 0.007}$ and
$\bar\eta = 0.343\, {}^{+\, 0.011}_{-\, 0.012}$,
where $\bar\rho\! -\! i \bar\eta\approx (1\! -\! \lambda^2 / 2)\, (\rho \! -\! i\eta)^{\phantom{l}}$  \cite{Charles:2015gya}.
This determination requires a good control of the strong interaction in flavour-changing transitions and makes use of many hadronic inputs obtained with lattice simulations or effective field theory tools \cite{ElKhadra}.

The significance of the global fit is shown in Fig.~\ref{fig:CKMfit}, which tests the unitarity relation $(V^\dagger V)_{bd}=0$, visualized as a triangle in the complex plane $(\bar\rho,\, \bar\eta)$ with its sides divided by $V^*_{cb} V^{\phantom{*}}_{cd}$. In the absence of CP violation ($\bar\eta=0$) the triangle would collapse into the real axis. All experimental constraints agree nicely, confirming the triangular structure and determining precisely its upper vertex $(\bar\rho,\, \bar\eta)$. The test can be further refined, considering only some subsets of data: tree-level transitions, loop processes, CP-conserving observables and signals of CP violation. All four subsets determine independently the same triangle, giving a solid consistency check of the unitarity relation and a highly non-trivial verification of the SM mechanism of CP violation.

The success of the SM description of flavour is deeply rooted in the unitarity structure of $V_{u_i d_j}$ and the associated GIM mechanism \cite{Glashow:1970gm}
which guarantees the absence of tree-level flavour-changing neutral currents (FCNCs). The subtle SM cancelations suppressing FCNC transitions would be easily destroyed in the presence of new physics contributions. Therefore,
flavour data provide very strong constraints on models with additional sources of flavour symmetry breaking and probe physics at energy scales not directly accessible at accelerators. For instance, an effective $\Delta B = 2$ interaction of the form
\bel{eq:DB=2}
\cL_{\Delta B = 2}\, =\, \frac{c^{\Delta B = 2}}{\Lambda_{\mathrm{NP}}^2}\; \left( b_L\gamma^\mu d_L\right) (b_L\gamma_\mu d_L)\, ,
\ee
induced by new physics at the scale $\Lambda_{\mathrm{NP}}$, is tightly constrained by the measured amount of $B^0$--$\bar B^0$ mixing:  $|c^{\Delta B = 2}/\Lambda_{\mathrm{NP}}^2| < 2.3\times 10^{-6}\;\mathrm{TeV}^{-2}$. A generic flavour structure with
$c^{\Delta B = 2}\sim\cO(1)$ is ruled out at the TeV scale. New physics at $\Lambda_{\mathrm{NP}}\sim 1$~TeV would only become possible if $c^{\Delta B = 2}$ inherits the strong SM suppressions induced by the GIM mechanism. This requirement
can be satisfied assuming that the up and down Yukawa matrices are the only sources of quark-flavour symmetry breaking (minimal flavour violation) \cite{Isidori}.

\subsection{Two-Higgs Doublet Models}
\label{subsec:2HDM}

The  non-generic nature of the flavour structure becomes apparent if one considers two scalar doublets $\Phi_a$, which increases the number of quark Yukawas:
\bel{eq:GenYukawa}
\cL_Y = -\sum_{a=1}^2\;\left\{ \bar Q_L'
\cY^{(a)}_d\Phi_a\, d_R'
+ \bar Q_L'\cY^{(a)}_u\tilde\Phi_a\, u_R' \right\}\, .
\ee
Here, all fermionic fields are 3-dimensional flavour vectors and
$\tilde\Phi_a \equiv i \tau_2\,{\Phi_a^*}$.
The flavour matrices $\cY^{(1)}_f$ and $\cY^{(2)}_f$ % ($f=u,d$)
are in general unrelated and cannot be diagonalized simultaneously, generating dangerous FCNCs. Unless the Yukawa couplings are very small or the scalar bosons very heavy, a very specific flavour structure is required \cite{Nebot}. The usually adopted solution imposes a discrete $Z_2$ symmetry to force one of the two Yukawa matrices to be zero \cite{Glashow:1976nt,Paschos:1976ay}; this leads to five different models with `natural flavour conservation' (types I, II, X, Y and inert) \cite{Wiebusch,Kikuchi}.
A more general possibility is to impose the alignment in flavour space of
$\cY^{(1)}_f$ and $\cY^{(2)}_f$ (proportional matrices) \cite{Pich:2009sp},
which eliminates FCNCs at tree level. Although quantum corrections generate a small misalignment, %of the Yukawas,
the flavour symmetries of the aligned two-Higgs doublet model (A2HDM) tightly constrain the possible FCNC structures, keeping their effects well below the present experimental bounds \cite{Pich:2009sp,Jung:2010ik}.

Flavour alignment results in a very specific structure, with all fermion-scalar interactions being proportional to the corresponding fermion masses. The Yukawas are fully characterized by three complex alignment parameters $\varsigma_f$ ($f=u,d,\ell$), which introduce new sources of CP violation. All $Z_2$ models are recovered for particular real values of these parameters ($\varsigma_f = \cot\beta$ or $\varsigma_f = -\tan\beta$).

The A2HDM contains five physical scalars ($h$, $H$, $A$ and $H^\pm$), leading to a rich collider phenomenology. Fig.~\ref{fig:A2HDMfit} shows the present constraints on the up-type and down-type quark Yukawa couplings of the lightest CP-even scalar $h$, assuming that it is the boson discovered at 125 GeV and neglecting CP-violating effects \cite{Celis:2013ixa}.

%%%%%%%%%%%%%%%%%%%%%%% A2HDM fit %%%%%%%%%%%%%%%%%%%%%%%%%%%%%%%%%%%%
\begin{figure}[t]
\centering
\includegraphics[width=\columnwidth,clip]{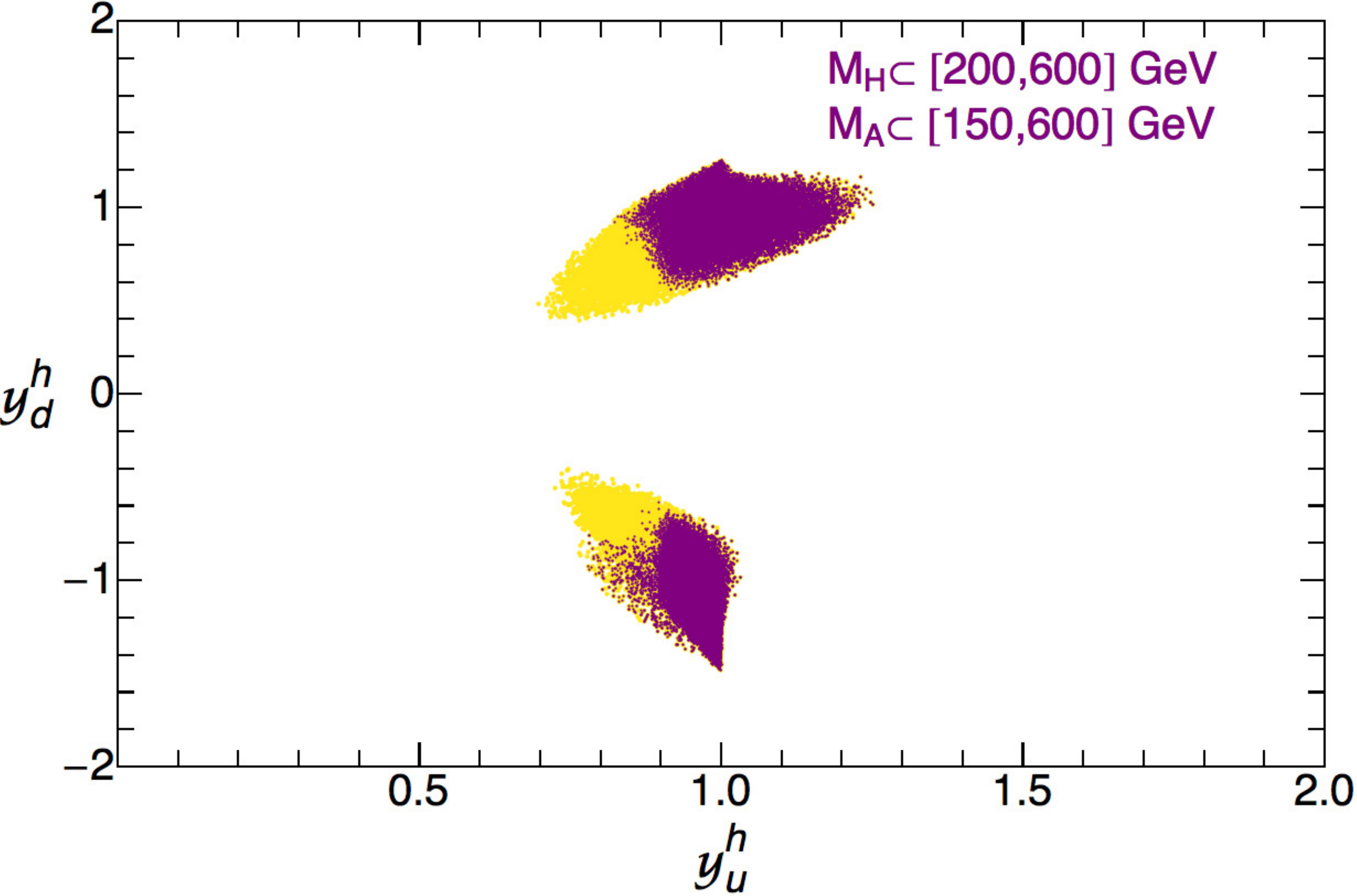} %7.5cm
\caption{Allowed 90\% CL regions (yellow) of the $y_q^h$ Yukawas (in SM units), in the CP-conserving A2HDM, from a global fit of collider and flavour data.
The purple regions include additional constraints from high-energy searches of the heavier neutral scalars $H$ and $A$ \cite{Celis:2013ixa}.} %,Celis:2013rcs}.}
\label{fig:A2HDMfit}
\end{figure}
%%%%%%%%%%%%%%%%%%%%%%%%%%%%%%%%%%%%%%%%%%%%%%%%%%%%%%%%%%%%%%%%%%%%

\subsection{Rare Decays}

The SM GIM suppression of loop-induced rare decays makes them a good testing ground for new-physics effects. The dimuon decays of the $B^0$ and $B^0_s$ mesons have been measured by CMS and LHCb \cite{CMS:2014xfa}:
\beqn\label{eq:Bmm}
\mathrm{Br}(B_s^0\to\mu^+\mu^-) & = & (2.8\,{}^{+\, 0.7}_{-\, 0.6})\times 10^{-9}\, ,
\no\\
\mathrm{Br}(B^0\to\mu^+\mu^-) & = & (3.9\,{}^{+\, 1.6}_{-\, 1.4})\times 10^{-10}\, ,
\eeqn
in good agreement with their predicted \cite{Bobeth:2013uxa} SM values $(3.65\pm 0.23)\times 10^{-9}$ and $(1.06\pm 0.09)\times 10^{-10}$, respectively. These rates are sensitive to new-physics contributions from extended scalar sectors \cite{Li:2014fea}.

A recent LHCb analysis of the angular distribution in $B^0\!\to\! K^{*0}(\to\!\! K\pi)\,\mu^+\mu^-$ \cite{Aaij:2013qta} finds a $3.7\,\sigma$ deviation from the SM \cite{Descotes-Genon:2014uoa,Altmannshofer:2014rta} for a particular optimized observable $P_5'$, representing the interference between the longitudinal and perpendicular $K^{*0}$ amplitudes, which has small hadronic uncertainties. The tension appears in Fig.~\ref{fig:P5}, for dilepton invariant masses between 4 and 8 $\mathrm{GeV}^2$, and suggests a sizeable negative shift in the Wilson coefficient of the effective operator
$O_9 = (\bar s_L\gamma_\mu  b_L)(\bar\ell\gamma^\mu\ell)$.
Another anomaly has shown up in $B^+\to K^+\ell^+\ell^-$, where the ratio between produced muons and electrons for dilepton invariant masses between 1 and 6
$\mathrm{GeV}^2$ is found to be $R_K = 0.745{}^{+0.090}_{-0.074}\pm 0.036$ \cite{Aaij:2014ora}, $2.6\,\sigma$ below the SM expectation \cite{Bobeth:2011nj}.

%%%%%%%%%%%%%%%%%%%%%%% A2HDM fit %%%%%%%%%%%%%%%%%%%%%%%%%%%%%%%%%%%%
\begin{figure}[t]
\centering
\includegraphics[width=\columnwidth,clip]{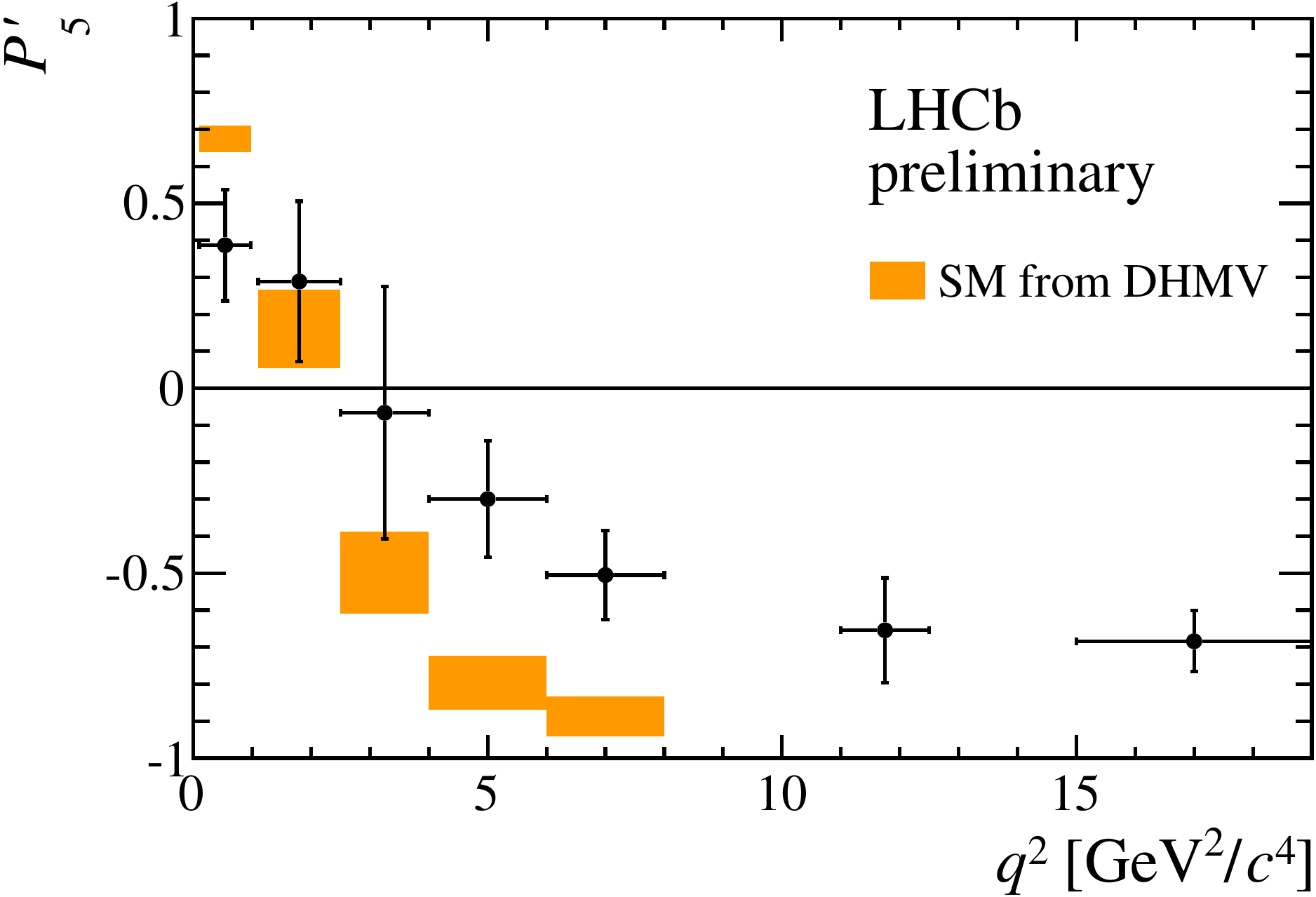} %7.5cm
\caption{Dilepton invariant-mass distribution of $P'_5$ \cite{Aaij:2013qta}.}
%The observable $P'_5$ in bins of the dilepton invariant mass \cite{Aaij:2013qta}.}
\label{fig:P5}
\end{figure}
%%%%%%%%%%%%%%%%%%%%%%%%%%%%%%%%%%%%%%%%%%%%%%%%%%%%%%%%%%%%%%%%%%%%

A different flavour anomaly has been found by BaBar in the tree-level decays $\bar B\to D^{(*)}\ell^-\bar\nu_\ell$ \cite{Lees:2012xj}, with a measured ratio between $\ell=\tau$ and $\ell=\mu, e$ events significantly higher than the SM prediction (2.0 and $2.7\,\sigma$ for $D$ and $D^*$) \cite{Fajfer:2012jt,Celis:2012dk}.
Another puzzling result is the like-sign dimuon charge asymmetry measured by D0 \cite{Abazov:2013uma}, which is $3.6\,\sigma$ above the expected SM prediction from $B^0_{d,s}$ mixing \cite{Bernardi}.
While more precise data is needed to clarify the situation, all these signals show the potential of flavour data to uncover new physics at higher scales.

\subsection{Lepton-Flavour Violation}

We have clear experimental evidence that neutrinos are massive particles and there is mixing in the lepton sector. The solar, atmospheric, accelerator and reactor neutrino
data lead to a consistent pattern of oscillation parameters \cite{Agashe:2014kda,Zito,Goswami,Tortola}.
The solar data determine $\Delta m^2_{21} \equiv m^2_2 - m^2_1 > 0$, but
the sign of $\Delta m^2_{31}$ is not yet established
and there can be two possibilities: normal ($m_1\!\! <\!\! m_2\!\! <\!\! m_3$) and
inverted ($m_3\!\! <\!\! m_1\!\! <\!\! m_2$) hierarchy.
The main recent advance is the determination of a sizeable non-zero value of $\theta_{13}$, confirming the 3$\nu$ mixing paradigm. A very precise value has been reported by Daya Bay:
$\sin^2{2\theta_{13}} = 0.084\pm 0.005$ \cite{An:2015rpe,Wang}.
This increases the interest for a next-generation of long-baseline neutrino experiments to measure the CP-violating phase $\delta_{\mathrm{CP}}$ and resolve the mass hierarchy.

The non-zero neutrino masses indicate new physics beyond the SM. Singlet $\nu_R$ fields are an obvious possibility, allowing for right-handed Majorana masses of arbitrary size which violate lepton number by two units. A very large Majorana mass scale can explain the tiny size of the observed neutrino masses through the well-known see-saw mechanism \cite{Minkowski:1977sc,GMRS:79,YA:79}. Nevertheless, an enlarged SM with 3 light $\nu_R$ fields has also been shown to be a viable phenomenological scenario \cite{Canetti:2012kh}.

Taking only into account the known $\nu_L$ fields, we can write the most general % Lagrangian invariant under
$\mathrm{SU(3)}_C\otimes \mathrm{SU(2)}_L\otimes \mathrm{U(1)}_Y$
invariant Lagrangian.
The SM is the unique answer with dimension four. The first contributions from new physics appear through dimension-5 operators, and have also a unique form which violates lepton number \cite{Weinberg:1979sa}:
\bel{eq:WE} \Delta \cL\; =\; - {c_{ij}\over\Lambda_{\mathrm{NP}}}\; \bar
L_i\,\tilde\Phi\, \tilde\Phi^t\, L_j^c \; + \; \mathrm{h.c.}\, , \ee
where $L_i$ denotes the $i$-flavoured lepton doublet and $L_i^c \equiv \mathcal{C} \bar L_i^T$. The EWSB generates a Majorana mass term for the left-handed neutrinos with\
$m_{ij} = c_{ij} v^2/\Lambda_{\mathrm{NP}}$. Taking $m_\nu\gtrsim 0.05$~eV, as suggested by atmospheric neutrino data, one gets
$\Lambda_{\mathrm{NP}}/c_{ij}\lesssim 10^{15}$~GeV, amazingly close to the
expected scale of Gran Unification.

With $m_{\nu_i}\not= 0$, the leptonic charged-current interactions involve a flavour mixing matrix $V_L$. The data on neutrino oscillations imply the following $3\,\sigma$ CL ranges for the magnitudes of the different $V_L$ entries \cite{Gonzalez-Garcia:2014bfa}:
\beqn %l{eq:V_L2}
\hskip -.65cm
|V_L| &\!\!\!\! = & \!\!\!\!
{\setlength{\arraycolsep}{4pt}
\left(
\begin{array}{ccc}
0.801\! -\! 0.845 & 0.514\! -\! 0.580 & 0.137\! -\! 0.158 \\
0.225\! -\! 0.517 & 0.441\! -\! 0.699 & 0.614\! -\! 0.793 \\
0.246\! -\! 0.529 & 0.464\! -\! 0.713 & 0.590\! -\! 0.776
\ea\right)\, .}
\nonumber\eeqn
Therefore, the mixing among leptons appears to be very different from the one in the quark sector.

The smallness of neutrino masses induces a strong suppression of neutrinoless
lepton-flavour-violating  transitions, which can be avoided in models with sources of lepton-flavour violation (LFV) not related to $m_{\nu_i}$.
LFV processes have the potential to probe physics at scales much higher than the TeV
\cite{Isidori}. The LFV scale can be constrained imposing the requirement of a viable leptogenesis. Recent studies within different new-physics scenarios find interesting correlations between $\mu$ and $\tau$ LFV decays, with $\mu\to e\gamma$ often expected to be close to the present exclusion limit.

CMS has obtained the first bound on LFV in Higgs decays: $\mathrm{Br}(H\to\mu^\pm\tau^\mp)<1.51\%$ (95\% CL) \cite{Khachatryan:2015kon}. This improves by one order of magnitude the indirect constraints on the corresponding Higgs Yukawas from $\tau$ decays \cite{Isidori,Celis:2014roa}.

\section{Searching for New Physics}
\label{sec:NewPhysics}

Although the SM could be valid up to arbitrary high scales, new dynamics should exist because we are lacking a proper understanding of important physical facts, such as the matter-antimatter asymmetry, the pattern of flavour mixings and fermion masses, the nature of dark matter or the accelerated expansion of the Universe.
However, the only signals of new physics detected so far are the non-zero neutrino masses and a few (not yet significant) flavour anomalies. All direct searches have resulted in negative results, pushing the new-physics scale beyond the reached experimental sensitivity \cite{Wuerthwein,Kim}.

\subsection{Desperately Seeking SUSY}

Neither the Tevatron nor the LHC have found any convincing evidence of supersymmetry (SUSY). Strong lower bounds on the masses of SUSY partners have been set, surpassing the TeV in many cases. Moreover, the Higgs mass creates a clear tension, being heavier than what was expected to be naturally accommodated in the minimal SUSY model (MSSM):
\beqn\label{eq:eps}
\mbox{}\hskip -.5cm
M_h^2 &\!\!\!\lesssim &\!\!\! M_Z^2 \cos^2{2\beta}
\\ &\!\!\! + &\!\!\!
\frac{3 m_t^4}{2\pi^2 v^2\sin^2{\beta}}\left[
\log{\left(\frac{M_{\tilde t}^2}{m_t^2}\right)} + \frac{X_t^2}{M_{\tilde t}^2}
\left(1-\frac{X_t^2}{12 M_{\tilde t}^2}\right)\right]\, ,
\no\eeqn
with $M_{\tilde t}^2\! =\! M_{\tilde t_1} M_{\tilde t_2}$.
% the geometric mean of the two stop masses. % and $X_t$ the stop mixing.
Masses as heavy as 125 GeV can only be reached in the decoupling regime ($M_{A,\tilde t}\gg M_Z$), with maximal stop mixing $X_t$ and $\tan{\beta} \ge 10$ \cite{Djouadi:2013lra}.
Although improved calculations including higher-order corrections allow for slightly larger values of $M_h$ \cite{Hahn:2013ria}, the situation looks bad in the usual constrained models (CMSSM, NUHM1, NUHM2, etc.) \cite{Bechtle:2014yna}, where the 120 MSSM parameters are reduced to just a few (4 plus 1 sign in CMSSM). A global fit to the data is still possible with heavy SUSY masses ($\gtrsim 1$~TeV), but only if the muon anomalous magnetic moment is not included in the fit, since the measured value of $(g\! -\! 2)_\mu$ \cite{Pich:2013lsa,Carlini} can no longer be explained \cite{Buchmueller:2014yva}.

With a larger set of 10 \cite{deVries:2015hva} or even 19--20 \cite{Cahill-Rowley:2014twa} free parameters, the Phenomenological MSSM (pMSSM) provides more flexibility and it is possible to find many parameter sets which are consistent with the data. In particular, compressing the SUSY spectrum, allows for still undetected light sparticles \cite{Cahill-Rowley:2014twa}.

Many SUSY variants (NMSSM, Split, High-Scale, Stealth, 5D, Natural, Folded, Twin, etc.) have been advocated to conform with the present experimental situation \cite{Carena}.
While some of them can be theoretically motivated (naturalness, dark matter, etc.), in most cases this is a data-driven search, looking for mechanisms to hide and avoid the strong data constraints. SUSY appears to be badly broken which questions its possible role in protecting the electroweak scale.

%%% \subsection{Exotica}
%%% \subsection{Cosmology and the Dark Side}
\subsection{Looking into the Dark Side}
% {Dark Matter. Hidden Portals}

Several astrophysical and cosmological evidences indicate that dark matter (DM)
is the dominant matter component in our Universe, accounting for 26.8\% of its
total energy budget \cite{Ade:2013zuv,Martinez}. Weakly interacting massive particles (WIMPs) around the TeV scale %, $\chi$,
are considered among the leading DM candidates, because they would have the right annihilation cross section (WIMP miracle) in the early Universe, after the thermal freeze-out, to explain the present DM relic density.
Very light axion-like particles are also an alternative DM possibility.
Sensitive searches for DM are performed by observing their annihilation (or decay) products ($\chi +\chi\to$ SM + SM)
% (DM + DM $\to$ SM + SM)
in satellite \cite{Berdugo} or $\gamma$-ray \cite{Stegmann} experiments, through their direct detection in earth-based detectors ($\chi +$ SM $\to\chi +$ SM)
%(DM + SM $\to$ DM + SM)
\cite{Araujo} or their production at high-energy colliders (SM + SM $\to\chi +\chi $)
% (SM + SM $\to$ DM + DM)
\cite{Wuerthwein}.

Viable (neutral, cold and stable) DM candidates exist in many models, especially those inhibiting their decay through some symmetry, such as the lightest SUSY (R parity) or little-Higgs T-odd particles, or a $Z_2$-odd scalar in the inert two-Higgs doublet model.
The experimental bounds on DM cross sections strongly constrain the parameter space of these models \cite{Carena,Sokolowska}.

DM could also be associated with a hidden sector, {\it i.e.} new particles that are singlets under the SM gauge group, which could be accessible through their couplings with SM singlet operators. For instance, the operator\ $\bar L_i \tilde\Phi$ could couple to new neutral singlet fermions (neutrino portal) \cite{Lindner}, and a new Abelian gauge field strength $F'_{\mu\nu}$ could be detected \cite{Palladino} through its mixing with the SM $U(1)_Y$ field (vector portal: $F'_{\mu\nu}F^{\mu\nu}_Y$). The square of the Higgs field, $\Phi^\dagger\Phi$, provides now a very interesting scalar portal to be explored, coupling either to new singlet scalars ($S$, $S^2$) or fermion bilinears ($\bar\psi\psi$).

%\subsection{Cosmology}
\subsection{Far-Away Messengers and Cosmology}

Ultrahigh-energy cosmic \cite{Berdugo,Matthews} and $\gamma$ \cite{Stegmann} rays, and neutrinos \cite{Kappes}, bring information from very distant astrophysical sources,
while cosmology data sheds light into the origin, evolution and large-scale structure of the Universe.
The most recent Planck data \cite{Ade:2013zuv}, based on the full-mission observations of the cosmic microwave background temperature and polarization anisotropies,
% lensing-potential power spectra,
are in excellent agreement with the standard % spatially-flat
6-parameter $\Lambda$CDM cosmology with a power-law spectrum of adiabatic scalar perturbations.
Our Universe appears to be spatially flat to an accuracy of 0.5\%  \cite{Ade:2013zuv}:
\bel{eq:Omega}
\Omega_{\mathrm{tot}}\, =\, 1.000\pm 0.005\,.
\ee

Cosmology provides also precious information on the absolute neutrino mass scale. The combination of Planck temperature data  with measurements of the baryon acoustic oscillation angular scale gives the 95\% CL upper bound \cite{Lindner,Lesgourgues:2014zoa}
\bel{eq:nu-mass}
\sum_i m_{\nu_i} \, <\, 0.23\:\mathrm{eV}\, .    %0.28 ?
\ee

A controversial highlight at ICHEP 2014 \cite{Martinez,Obrient} was the BICEP2 observation of a large B-mode polarization anisotropy, which was suggested to be a signal of inflationary gravitational waves \cite{Ade:2014xna}. A detailed BICEP2-Planck joint analysis \cite{Ade:2015tva} has shown later that foreground polarized dust emission is responsible for a large part of the BICEP2 signal, finding no statistically significant evidence for tensor modes. The most recent Planck analysis gives a tight upper limit on the tensor-to-scalar ratio: $r_{0.002} < 0.09$ (95\% CL) \cite{Ade:2013zuv}.

The BICEP2 claim has triggered many recent studies of the inflationary paradigm \cite{Guth,Barenboim:2014tca}. While the excitement has now cooled down, it is worth stressing that models of inflation are usually based on scalar fields. The `criticality' of the SM scalar potential (with the measured Higgs and top masses) has reopened the viability of the most economical Higgs-inflation scenario, where the Higgs field plays the inflaton role \cite{Bezrukov:2014bra,Hamada:2014iga}.

\section{Electroweak Effective Field Theory}

The non-observation of new states suggests the existence of a mass gap between the electroweak and new-physics scales. This can be adequately described with effective field theory methods \cite{Pich:1998xt}, writing the most general Lagrangian with the SM gauge symmetries in terms of the known light ($m\!\ll\!\Lambda_{\mathrm{NP}}$) fields \cite{Freitas}:
\bel{eq:EFT}
\cL_{\mathrm{eff}}\, =\, \cL_{\mathrm{SM}} + \sum_{D>4}\sum_k \frac{c_k^{(D)}}{\Lambda_{\mathrm{NP}}^{D-4}}\, O_k^{(D)}\, .
\ee
The lowest-order term is the dimension-4 SM, while low-energy signals of new physics are parametrized in terms of higher-dimensional operators suppressed by the corresponding powers of the new-physics scale. The information on the unknown underlying dynamics is then contained in the coefficients $c_k^{(D)}$.

The lepton-number violating operator \eqn{eq:WE} is the only possibility with $D=5$.
With a single SM family, there are 59 (5) independent operators with $D\! =\! 6$ which preserve (violate) B and L \cite{Grzadkowski:2010es,Buchmuller:1985jz}. Although the present data are not good enough to perform a global analysis with the full set of operators, interesting constraints on restricted subsets start to emerge, specially those accessible through electroweak precision tests and Higgs data \cite{Falkowski:2014tna}.
Unfortunately, the number of operators blows up to 2499 (assuming B and L conservation) when 3-generation flavour quantum numbers are included \cite{Alonso:2013hga}. Hints on the flavour structure are clearly needed.

The previous operator counting assumes the SM Higgs doublet structure. A more general approach, also valid in strongly-coupled scenarios of EWSB, builds the effective Lagrangian with a singlet Higgs field, using only the known global symmetries of the EWSB.
The SM scalar Lagrangian \eqn{eq:Lphi} can be rewritten as \cite{Pich:1998xt}
\bel{eq:Lphi2}
\cL_\Phi\, =\, \frac{v^2}{4}\,\,\mathrm{Tr}\left[ \left(D^\mu U\right)^\dagger D_\mu U\right]\, +\, \cO(H/v)\, ,
\ee
where $U\! =\! \exp{(i\vec\sigma\vec\phi/v)}$
% $U\! =\! \exp{(\frac{i}{v}\vec\sigma\vec\phi)}$
parametrizes the Goldstone bosons, making manifest the pattern of `custodial' symmetry breaking: $\mathrm{SU(2)}_L\otimes \mathrm{SU(2)}_R\to \mathrm{SU(2)}_V$ ($\cL_\Phi$ is invariant under
$U\to g_L^{\phantom{\dagger}} U g_R^\dagger$ with $g_{L,R}^{\phantom{\dagger}}$ arbitrary global $\mathrm{SU(2)}_{L,R}$ transformations). The same Goldstone Lagrangian describes pion dynamics in QCD, with the changes $v\to f_\pi$ and $\vec\phi\to\vec\pi$.

The Goldstone covariant derivatives generate the $W^\pm$ and $Z$ masses, which are not necessarily related to the Higgs field. Since the electroweak Goldstones constitute the longitudinal polarizations of the gauge bosons, the scattering $V_L V_L\to V_L V_L$
($V=W^\pm, Z$) directly tests the Goldstone dynamics. With generic $HV^2$ and $H^2V^2$ couplings, the tree-level scattering matrix grows as $s/v^2$ at high energies, while one-loop corrections induce a much worse ultraviolet behaviour $s^2\log{s}/v^4$ \cite{Delgado:2013hxa,Espriu:2013fia}. This entails a violation of unitarity, which is only canceled if the gauge couplings of the Higgs boson take exactly their SM values. Any small deviation in the Higgs couplings would necessarily imply the presence of new-physics contributions to $V_L V_L$ scattering in order to restore unitarity.
Therefore, the measurement of $\sigma(V_L V_L\to V_L V_L)$ at the LHC is a very important, but difficult, challenge. The first evidence of $W^\pm W^\pm$ collisions has been recently reported by ATLAS \cite{Aad:2014zda}, and used to set limits on anomalous quartic gauge couplings.

\section{Status and Outlook}

After the Higgs discovery, the SM framework is now fully established as the correct theory at the electroweak scale. It successfully explains the experimental results with high precision and all its ingredients have been verified. With the measured Higgs and top masses, the SM could even be a valid theory up to the Planck scale.

However, new physics is still needed to explain many pending questions for which the SM does not provide satisfactory answers. A proper understanding of the vastly different mass scales spanned by the known particles is missing. The dynamics of flavour and the origin of CP violation are also related to the mass generation. The Higgs boson could well be a window into unknown dynamical territory. Thus, its properties must be analyzed with high precision to uncover any possible deviation from the SM. The present data are already putting stringent constraints on alternative scenarios of EWSB and pushing the scale of new physics to higher energies. How far this scale could be is an open question of obvious experimental relevance.

The LHC data are challenging our previous ideas about naturalness and the TeV scale.
The most fashionable new-physics scenarios are now less compelling than before,  making us suspect that Nature has chosen a quite different path.
The upcoming LHC run could bring unexpected surprises, changing our views on fundamental physics. We are awaiting for great discoveries; the LHC scientific adventure is just starting.

\section*{Acknowledgements}
I would like to thank all ICHEP 2014 participants for their many interesting contributions, and the organizers for inviting me to present this summary.
This work has been supported by the Spanish
Government and ERDF funds from the EU Commission [Grants
No. FPA2011-23778, FPA2014-53631-C2-1-P and CSD2007-
00042 (Consolider Project CPAN)] and by Generalitat
Valenciana under Grant No. PrometeoII/2013/007.

%% The Appendices part is started with the command \appendix;
%% appendix sections are then done as normal sections
%% \appendix

%% \section{}
%% \label{}

%% References
%%
%% Following citation commands can be used in the body text:
%% Usage of \cite is as follows:
%%   \cite{key}         ==>>  [#]
%%   \cite[chap. 2]{key} ==>> [#, chap. 2]
%%

%% References with BibTeX database:
\nocite{*}
\bibliographystyle{elsarticle-num}
%\bibliography{martin}

\begin{thebibliography}{00}

%\cite{Aad:2012tfa}
\bibitem{Aad:2012tfa} ATLAS Collaboration,
 % G.~Aad {\it et al.}  [ATLAS Collaboration],
  %``Observation of a new particle in the search for the Standard Model Higgs boson with the ATLAS detector at the LHC,''
  Phys.\ Lett.\ B {\bf 716} (2012) 1.
 % [arXiv:1207.7214 [hep-ex]].
  %%CITATION = ARXIV:1207.7214;%%
  %3887 citations counted in INSPIRE as of 24 Jan 2015

%\cite{Chatrchyan:2012ufa}
\bibitem{Chatrchyan:2012ufa} CMS Collaboration,
 % S.~Chatrchyan {\it et al.}  [CMS Collaboration],
  %``Observation of a new boson at a mass of 125 GeV with the CMS experiment at the LHC,''
  Phys.\ Lett.\ B {\bf 716} (2012) 30.
 % [arXiv:1207.7235 [hep-ex]].
  %%CITATION = ARXIV:1207.7235;%%
  %3816 citations counted in INSPIRE as of 24 Jan 2015

\bibitem{Nason} P. Nason, ICHEP 2014. %% these proceedings.

\bibitem{Uwer} P. Uwer, ICHEP 2014. %% these proceedings.

%\cite{Anastasiou:2015ema}
\bibitem{Anastasiou:2015ema} C.~Anastasiou {\it et al.},
  % C.~Anastasiou, C.~Duhr, F.~Dulat, F.~Herzog and B.~Mistlberger,
  %``Higgs boson gluon-fusion production in N3LO QCD,''
%  arXiv:1503.06056 [hep-ph].
  %%CITATION = ARXIV:1503.06056;%%
  %3 citations counted in INSPIRE as of 24 Apr 2015
%\cite{Anastasiou:2015vya}
%\bibitem{Anastasiou:2015vya}
%  C.~Anastasiou, C.~Duhr, F.~Dulat, F.~Herzog and B.~Mistlberger,
  %``Higgs Boson Gluon-Fusion Production in QCD at Three Loops,''
  Phys.\ Rev.\ Lett.\  {\bf 114} (2015) 21,  212001.
%  [arXiv:1503.06056 [hep-ph]].
  %%CITATION = ARXIV:1503.06056;%%
  %13 citations counted in INSPIRE as of 04 juin 2015

\bibitem{Roda} C. Roda, ICHEP 2014. %% these proceedings.

\bibitem{Carli} T. Carli, ICHEP 2014. %%  these proceedings.

\bibitem{Berryhill} J. Berryhill, ICHEP 2014. %% these proceedings.


%\cite{Pich:2013lsa}
\bibitem{Pich:2013lsa}
  A.~Pich,
  %``Precision Tau Physics,''
  Prog.\ Part.\ Nucl.\ Phys.\  {\bf 75} (2014) 41.
 % [arXiv:1310.7922 [hep-ph]].
  %%CITATION = ARXIV:1310.7922;%%
  %23 citations counted in INSPIRE as of 24 gen 2015

%\cite{Malaescu:2012ts}
\bibitem{Malaescu:2012ts}
  B.~Malaescu and P.~Starovoitov,
  %``Evaluation of the Strong Coupling Constant alpha_s Using the ATLAS Inclusive Jet Cross-Section Data,''
  Eur.\ Phys.\ J.\ C {\bf 72} (2012) 2041.
 % [arXiv:1203.5416 [hep-ph]].
  %%CITATION = ARXIV:1203.5416;%%
  %25 citations counted in INSPIRE as of 24 Apr 2015

%\cite{Chatrchyan:2013txa}
\bibitem{Chatrchyan:2013txa} CMS Collaboration,
  %S.~Chatrchyan {\it et al.}  [CMS Collaboration],
  %``Measurement of the ratio of the inclusive 3-jet cross section to the inclusive 2-jet cross section in pp collisions at $\sqrt{s}$ = 7 TeV and first determination of the strong coupling constant in the TeV range,''
  Eur.\ Phys.\ J.\ C {\bf 73} (2013) 10, 2604;
  % [arXiv:1304.7498 [hep-ex]].
  %%CITATION = ARXIV:1304.7498;%%
  %38 citations counted in INSPIRE as of 31 Mar 2015
%
%\cite{CMS:2014mna}
%\bibitem{CMS:2014mna}
%  V.~Khachatryan {\it et al.}  [CMS Collaboration],
  %``Measurement of the inclusive 3-jet production differential cross section in proton–proton collisions at 7 TeV and determination of the strong coupling constant in the TeV range,''
 % Eur.\ Phys.\ J.\
 C {\bf 75} (2015) 5,  186;
 % [arXiv:1412.1633 [hep-ex]].
  %%CITATION = ARXIV:1412.1633;%%
  %5 citations counted in INSPIRE as of 24 Apr 2015
%
%\cite{Chatrchyan:2013haa}
%\bibitem{Chatrchyan:2013haa} CMS Collaboration,
  % S.~Chatrchyan {\it et al.}  [CMS Collaboration],
  %``Determination of the top-quark pole mass and strong coupling constant from the t t-bar production cross section in pp collisions at $\sqrt{s}$ = 7 TeV,''
  Phys.\ Lett.\ B {\bf 728} (2014) 496 %, 526.
   [Erratum-ibid.\ B {\bf 728} (2014) 526];
 % [arXiv:1307.1907 [hep-ex]].
  %%CITATION = ARXIV:1307.1907;%%
  %56 citations counted in INSPIRE as of 31 Mar 2015
%
%\cite{Khachatryan:2014waa}
%\bibitem{Khachatryan:2014waa}
%  V.~Khachatryan {\it et al.}  [CMS Collaboration],
  %``Constraints on parton distribution functions and extraction of the strong coupling constant from the inclusive jet cross section in pp collisions at $\sqrt{s}$ = 7 TeV,''
  arXiv:1410.6765 [hep-ex].
  %%CITATION = ARXIV:1410.6765;%%
  %9 citations counted in INSPIRE as of 24 Apr 2015



%\cite{Agashe:2014kda}
\bibitem{Agashe:2014kda} Particle Data Group,
  % K.A.~Olive {\it et al.}  [Particle Data Group Collaboration],
  %``Review of Particle Physics,''
  Chin.\ Phys.\ C {\bf 38} (2014) 090001.
  %%CITATION = CHPHD,C38,090001;%%
  %491 citations counted in INSPIRE as of 23 Jan 2015

%\cite{Pich:2013sqa}
\bibitem{Pich:2013sqa}
  A.~Pich,
  %``Review of $\alpha_s$ determinations,''
  PoS ConfinementX {\bf } (2012) 022.
  %[arXiv:1303.2262 [hep-ph]].
  %%CITATION = ARXIV:1303.2262;%%
  %24 citations counted in INSPIRE as of 23 Jan 2015

\bibitem{ElKhadra} A. El-Khadra, ICHEP 2014. %% these proceedings.

%\cite{Aoki:2013ldr}
\bibitem{Aoki:2013ldr} FLAG Working Group,
  %S.~Aoki, Y.~Aoki, C.~Bernard, T.~Blum, G.~Colangelo, M.~Della Morte, S.~Dürr and A.~X.~El Khadra {\it et al.},
  %``Review of lattice results concerning low-energy particle physics,''
  Eur.\ Phys.\ J.\ C {\bf 74} (2014) 9,  2890.
%  [arXiv:1310.8555 [hep-lat]].
  %%CITATION = ARXIV:1310.8555;%%
  %153 citations counted in INSPIRE as of 23 Jan 2015

%\cite{Brambilla:2014jmp}
\bibitem{Brambilla:2014jmp} N.~Brambilla {\it et al.},
 % N.~Brambilla, S.~Eidelman, P.~Foka, S.~Gardner, A.~S.~Kronfeld, M.~G.~Alford, R.~Alkofer and M.~Butenschoen {\it et al.},
  %``QCD and Strongly Coupled Gauge Theories: Challenges and Perspectives,''
  Eur.\ Phys.\ J.\ C {\bf 74} (2014) 10,  2981.
 % [arXiv:1404.3723 [hep-ph]].
  %%CITATION = ARXIV:1404.3723;%%
  %65 citations counted in INSPIRE as of 24 Apr 2015

\bibitem{Peng} H. Peng, ICHEP 2014. %% these proceedings.

\bibitem{Wessels} J.P. Wessels, ICHEP 2014. %% these proceedings.

\bibitem{Gale} C. Gale, ICHEP 2014. %% these proceedings.

\bibitem{Barbon} J.L.F. Barb\'on, ICHEP 2014. %% these proceedings.


%\cite{Aaltonen:2015cra}
\bibitem{Aaltonen:2015cra} CDF and D0 Collaborations,
  % T.~A.~Aaltonen {\it et al.}  [CDF and D0 Collaborations],
  %``Tevatron combination of single-top-quark cross sections and determination of the magnitude of the Cabibbo-Kobayashi-Maskawa matrix element $\bf V_{tb}$,''
  arXiv:1503.05027 [hep-ex].
  %%CITATION = ARXIV:1503.05027;%%

%\cite{Khachatryan:2014iya}
\bibitem{Khachatryan:2014iya} CMS Collaboration,
  %V.~Khachatryan {\it et al.}  [CMS Collaboration],
  %``Measurement of the t-channel single-top-quark production cross section and of the $\mid V_{tb} \mid$ CKM matrix element in pp collisions at $\sqrt{s}$= 8 TeV,''
  JHEP {\bf 1406} (2014) 090.
%  [arXiv:1403.7366 [hep-ex]].
  %%CITATION = ARXIV:1403.7366;%%
  %25 citations counted in INSPIRE as of 24 gen 2015

%\cite{Kuhn:2011ri}
\bibitem{Kuhn:2011ri}
  J.H.~Kuhn and G.~Rodrigo,
  %``Charge asymmetries of top quarks at hadron colliders revisited,''
  JHEP {\bf 1201} (2012) 063.
 % [arXiv:1109.6830 [hep-ph]].
  %%CITATION = ARXIV:1109.6830;%%
  %146 citations counted in INSPIRE as of 24 Jan 2015

%\cite{Abazov:2014cca}
\bibitem{Abazov:2014cca} D0 Collaboration,
 % V.~M.~Abazov {\it et al.}  [D0 Collaboration],
  %``Measurement of the forward-backward asymmetry in top quark-antiquark production in ppbar collisions using the lepton+jets channel,''
  Phys.\ Rev.\ D {\bf 90} (2014) 7,  072011.
%  [arXiv:1405.0421 [hep-ex]].
  %%CITATION = ARXIV:1405.0421;%%
  %26 citations counted in INSPIRE as of 24 Jan 2015

\bibitem{Bernardi} G. Bernardi, ICHEP 2014. %%  these proceedings.

\bibitem{Aldaya} M. Aldaya, ICHEP 2014. %%  these proceedings.

%\cite{ATLAS:2014wva}
\bibitem{ATLAS:2014wva}
  ATLAS, CDF, CMS and D0 Collaborations,
  %``First combination of Tevatron and LHC measurements of the top-quark mass,''
  arXiv:1403.4427 [hep-ex].
  %%CITATION = ARXIV:1403.4427;%%
  %160 citations counted in INSPIRE as of 31 Mar 2015


%\cite{Hoang:2008xm}
\bibitem{Hoang:2008xm}
  A.H.~Hoang and I.W.~Stewart,
  %``Top Mass Measurements from Jets and the Tevatron Top-Quark Mass,''
  Nucl.\ Phys.\ Proc.\ Suppl.\  {\bf 185} (2008) 220.
%  [arXiv:0808.0222 [hep-ph]].
  %%CITATION = ARXIV:0808.0222;%%
  %66 citations counted in INSPIRE as of 24 gen 2015

%\cite{Moch:2014tta}
\bibitem{Moch:2014tta} S.~Moch {\it et al.},
 % S.~Moch, S.~Weinzierl, S.~Alekhin, J.~Blumlein, L.~de la Cruz, S.~Dittmaier, M.~Dowling and J.~Erler {\it et al.},
  %``High precision fundamental constants at the TeV scale,''
  arXiv:1405.4781 [hep-ph].
  %%CITATION = ARXIV:1405.4781;%%
  %32 citations counted in INSPIRE as of 24 gen 2015

%\cite{Aad:2014kva}
\bibitem{Aad:2014kva}  ATLAS Collaboration,
  % G.~Aad {\it et al.}  [ATLAS Collaboration],
  %``Measurement of the $t\overline{t}$ production cross-section using $e\mu $ events with $b$ -tagged jets in $pp$ collisions at $\sqrt{s}=7$ and 8 TeV with the ATLAS detector,''
  Eur.\ Phys.\ J.\ C {\bf 74} (2014) 10,  3109.
 % [arXiv:1406.5375 [hep-ex]].
  %%CITATION = ARXIV:1406.5375;%%
  %34 citations counted in INSPIRE as of 24 Jan 2015

%\cite{Alioli:2013mxa}
\bibitem{Alioli:2013mxa} S.~Alioli {\it et al.},
 % S.~Alioli, P.~Fernandez, J.~Fuster, A.~Irles, S.~O.~Moch, P.~Uwer and M.~Vos,
  %``A new observable to measure the top-quark mass at hadron colliders,''
  Eur.\ Phys.\ J.\ C {\bf 73} (2013) 2438.
 % [arXiv:1303.6415 [hep-ph]].
  %%CITATION = ARXIV:1303.6415;%%
  %23 citations counted in INSPIRE as of 24 gen 2015

%\cite{}
\bibitem{ATLAS-mt}
  ATLAS Collaboration,
  %``Determination of the top-quark pole mass using tt t+1-jet events collected with the ATLAS experiment in 7 TeV pp collisions,''
  ATLAS-CONF-2014-053.  %%%, ATLAS-COM-CONF-2014-069.
  %%CITATION = ATLAS-CONF-2014-053, ATLAS-COM-CONF-2014-069;%%
  %3 citations counted in INSPIRE as of 31 Mar 2015

%\cite{Kawabataa:2014osa}
\bibitem{Kawabataa:2014osa} S.~Kawabata {\it et al.},
 % S.~Kawabata, Y.~Shimizu, Y.~Sumino and H.~Yokoya,
  %``Weight function method for precise determination of top quark mass at Large Hadron Collider,''
  Phys.\ Lett.\ B {\bf 741} (2014) 232.
%  [arXiv:1405.2395 [hep-ph]].
  %%CITATION = ARXIV:1405.2395;%%
  %4 citations counted in INSPIRE as of 24 Jan 2015

\bibitem{Kado} M.M. Kado, ICHEP 2014. %%  these proceedings.

\bibitem{David} A. David, ICHEP 2014. %%  these proceedings.

%\cite{Aad:2015zhl}
\bibitem{Aad:2015zhl} ATLAS and CMS Collaborations,
  % G.~Aad {\it et al.}  [ATLAS and CMS Collaborations],
  %``Combined Measurement of the Higgs Boson Mass in $pp$ Collisions at $\sqrt{s}=7$ and 8 TeV with the ATLAS and CMS Experiments,''
  Phys.\ Rev.\ Lett.\  {\bf 114} (2015) 191803.
%  [arXiv:1503.07589 [hep-ex]].
  %%CITATION = ARXIV:1503.07589;%%
  %48 citations counted in INSPIRE as of 04 juin 2015

\bibitem{Grojean} C. Grojean, ICHEP 2014. %%  these proceedings.

%\cite{Higgs:1964pj}
\bibitem{Higgs:1964pj}
  P.W.~Higgs,
  %``Broken Symmetries and the Masses of Gauge Bosons,''
  Phys.\ Rev.\ Lett.\  {\bf 13} (1964) 508;
  %%CITATION = PRLTA,13,508;%%
  %2310 citations counted in INSPIRE as of 24 Jul 2013
%
%\cite{Higgs:1964ia}
%\bibitem{Higgs:1964ia}
%  P.~W.~Higgs,
  %``Broken symmetries, massless particles and gauge fields,''
  Phys.\ Lett.\  {\bf 12} (1964) 132;
  %%CITATION = PHLTA,12,132;%%
  %2486 citations counted in INSPIRE as of 24 Jul 2013
%
%\cite{Higgs:1966ev}
%\bibitem{Higgs:1966ev}
%  P.~W.~Higgs,
  %``Spontaneous Symmetry Breakdown without Massless Bosons,''
  Phys.\ Rev.\  {\bf 145} (1966) 1156.
  %%CITATION = PHRVA,145,1156;%%
  %1840 citations counted in INSPIRE as of 24 Jul 2013

%\cite{Englert:1964et}
\bibitem{Englert:1964et}
  F.~Englert and R.~Brout,
  %``Broken Symmetry and the Mass of Gauge Vector Mesons,''
  Phys.\ Rev.\ Lett.\  {\bf 13} (1964) 321.
  %%CITATION = PRLTA,13,321;%%
  %2100 citations counted in INSPIRE as of 24 Jul 2013

%\cite{Guralnik:1964eu}
\bibitem{Guralnik:1964eu}
  G.S.~Guralnik, C.R.~Hagen and T.W.B.~Kibble,
  %``Global Conservation Laws and Massless Particles,''
  Phys.\ Rev.\ Lett.\  {\bf 13} (1964) 585.
  %%CITATION = PRLTA,13,585;%%
  %1675 citations counted in INSPIRE as of 24 Jul 2013

%\cite{Kibble:1967sv}
\bibitem{Kibble:1967sv}
  T.W.B.~Kibble,
  %``Symmetry breaking in nonAbelian gauge theories,''
  Phys.\ Rev.\  {\bf 155} (1967) 1554.
  %%CITATION = PHRVA,155,1554;%%
  %926 citations counted in INSPIRE as of 24 Jul 2013

%\cite{Baak:2014ora}
\bibitem{Baak:2014ora}
  M.~Baak {\it et al.}, %  [Gfitter Group Collaboration],
  %``The global electroweak fit at NNLO and prospects for the LHC and ILC,''
  Eur.\ Phys.\ J.\ C {\bf 74} (2014) 9,  3046.
 % [arXiv:1407.3792 [hep-ph]].
  %%CITATION = ARXIV:1407.3792;%%
  %33 citations counted in INSPIRE as of 27 gen 2015

\bibitem{Freitas} A. Freitas, ICHEP 2014. %%  these proceedings.

%\cite{Buttazzo:2013uya}
\bibitem{Buttazzo:2013uya}
  D.~Buttazzo {\it et al.},
  % G.~Degrassi, P.~P.~Giardino, G.~F.~Giudice, F.~Sala, A.~Salvio and A.~Strumia,
  %``Investigating the near-criticality of the Higgs boson,''
  JHEP {\bf 1312} (2013) 089.
  % [arXiv:1307.3536 [hep-ph]].
  %%CITATION = ARXIV:1307.3536;%%
  %227 citations counted in INSPIRE as of 14 Feb 2015

%\cite{Alekhin:2012py}
\bibitem{Alekhin:2012py} S.~Alekhin {\it et al.},
  % S.~Alekhin, A.~Djouadi and S.~Moch,
  %``The top quark and Higgs boson masses and the stability of the electroweak vacuum,''
  Phys.\ Lett.\ B {\bf 716} (2012) 214.
  % [arXiv:1207.0980 [hep-ph]].
  %%CITATION = ARXIV:1207.0980;%%
  %161 citations counted in INSPIRE as of 14 Feb 2015

%\cite{Ilisie:2012cc}
\bibitem{Ilisie:2012cc}
  V.~Ilisie and A.~Pich,
  %``QCD exotics versus a Standard Model Higgs,''
  Phys.\ Rev.\ D {\bf 86} (2012) 033001.
 % [arXiv:1202.3420 [hep-ph]].
  %%CITATION = ARXIV:1202.3420;%%
  %8 citations counted in INSPIRE as of 14 Feb 2015

%\cite{Khachatryan:2014jba}
\bibitem{Khachatryan:2014jba} CMS Collaboration,
  % V.~Khachatryan {\it et al.}  [CMS Collaboration],
  %``Precise determination of the mass of the Higgs boson and tests of compatibility of its couplings with the standard model predictions using proton collisions at 7 and 8 $\,\text {TeV}$,''
  Eur.\ Phys.\ J.\ C {\bf 75} (2015) 5,  212.
%  [arXiv:1412.8662 [hep-ex]].
  %%CITATION = ARXIV:1412.8662;%%
  %102 citations counted in INSPIRE as of 04 Jun 2015

%\cite{Aad:2014eha}
\bibitem{Aad:2014eha} ATLAS Collaboration,
 % G.~Aad {\it et al.}  [ATLAS Collaboration],
  %``Measurement of Higgs boson production in the diphoton decay channel in pp collisions at center-of-mass energies of 7 and 8 TeV with the ATLAS detector,''
  Phys.\ Rev.\ D {\bf 90} (2014) 11,  112015.
%  [arXiv:1408.7084 [hep-ex]].
  %%CITATION = ARXIV:1408.7084;%%
  %54 citations counted in INSPIRE as of 14 Feb 2015

%%\cite{Aad:2013wqa}
%\bibitem{Aad:2013wqa} ATLAS Collaboration,
% % G.~Aad {\it et al.}  [ATLAS Collaboration],
%  %``Measurements of Higgs boson production and couplings in diboson final states with the ATLAS detector at the LHC,''
%  Phys.\ Lett.\ B {\bf 726} (2013) 88
%   [Erratum-ibid.\ B {\bf 734} (2014) 406].
% % [arXiv:1307.1427 [hep-ex]].
%  %%CITATION = ARXIV:1307.1427;%%
%  %327 citations counted in INSPIRE as of 14 Feb 2015

%\cite{Aad:2014eva}
\bibitem{Aad:2014eva} ATLAS Collaboration,
 % G.~Aad {\it et al.}  [ATLAS Collaboration],
  %``Measurements of Higgs boson production and couplings in the four-lepton channel in pp collisions at center-of-mass energies of 7 and 8 TeV with the ATLAS detector,''
  Phys.\ Rev.\ D {\bf 91} (2015) 1,  012006.
 % [arXiv:1408.5191 [hep-ex]].
  %%CITATION = ARXIV:1408.5191;%%
  %25 citations counted in INSPIRE as of 14 Feb 2015

 %\cite{ATLAS:2014aga}
\bibitem{ATLAS:2014aga} ATLAS Collaboration,
%  G.~Aad {\it et al.}  [ATLAS Collaboration],
  %``Observation and measurement of Higgs boson decays to $WW^{\ast}$ with the ATLAS detector,''
  arXiv:1412.2641 [hep-ex].
  %%CITATION = ARXIV:1412.2641;%%
  %3 citations counted in INSPIRE as of 14 Feb 2015


%\cite{Aad:2014xzb}
\bibitem{Aad:2014xzb} ATLAS Collaboration,
 % G.~Aad {\it et al.}  [ATLAS Collaboration],
  %``Search for the $b\bar{b}$ decay of the Standard Model Higgs boson in associated $(W/Z)H$ production with the ATLAS detector,''
  JHEP {\bf 1501} (2015) 069.
 % [arXiv:1409.6212 [hep-ex]].
  %%CITATION = ARXIV:1409.6212;%%
  %18 citations counted in INSPIRE as of 14 Feb 2015

%\cite{Aad:2015vsa}
\bibitem{Aad:2015vsa} ATLAS Collaboration,
 % G.~Aad {\it et al.}  [ATLAS Collaboration],
  %``Evidence for the Higgs-boson Yukawa coupling to tau leptons with the ATLAS detector,''
  JHEP {\bf 1504} (2015) 117.
  %[arXiv:1501.04943 [hep-ex]].
  %%CITATION = ARXIV:1501.04943;%%
  %17 citations counted in INSPIRE as of 24 Apr 2015

%\cite{Ellis:2013lra}
\bibitem{Ellis:2013lra}
  J.~Ellis and T.~You,
  %``Updated Global Analysis of Higgs Couplings,''
  JHEP {\bf 1306} (2013) 103.
%  [arXiv:1303.3879 [hep-ph]].
  %%CITATION = ARXIV:1303.3879;%%
  %144 citations counted in INSPIRE as of 14 Feb 2015

\bibitem{Browder} T.E. Browder, ICHEP 2014. %%  these proceedings

\bibitem{Albrecht} J. Albrecht, ICHEP 2014. %%  these proceedings

%\cite{Wolfenstein:1983yz}
\bibitem{Wolfenstein:1983yz}
  L.~Wolfenstein,
  %``Parametrization of the Kobayashi-Maskawa Matrix,''
  Phys.\ Rev.\ Lett.\  {\bf 51} (1983) 1945.
  %%CITATION = PRLTA,51,1945;%%
  %2503 citations counted in INSPIRE as of 14 Feb 2015

%\cite{Charles:2015gya}
\bibitem{Charles:2015gya} J.~Charles {\it et al.},
  %J.~Charles, O.~Deschamps, S.~Descotes-Genon, H.~Lacker, A.~Menzel, S.~Monteil, V.~Niess and J.~Ocariz {\it et al.},
  %``Current status of the Standard Model CKM fit and constraints on $\Delta F=2$ New Physics,''
  Phys.\ Rev.\ D {\bf 91} (2015) 7,  073007.
 % [arXiv:1501.05013 [hep-ph]].
  %%CITATION = ARXIV:1501.05013;%%
  %10 citations counted in INSPIRE as of 24 Apr 2015

%\cite{Glashow:1970gm}
\bibitem{Glashow:1970gm} S.L.~Glashow {\it et al.},
  % S.L.~Glashow, J.~Iliopoulos and L.~Maiani,
  %``Weak Interactions with Lepton-Hadron Symmetry,''
  Phys.\ Rev.\ D {\bf 2} (1970) 1285.
  %%CITATION = PHRVA,D2,1285;%%
  %4837 citations counted in INSPIRE as of 24 Apr 2015

\bibitem{Isidori} G. Isidori, ICHEP 2014. %%  these proceedings.

\bibitem{Nebot} M. Nebot, ICHEP 2014. %%  these proceedings.

%\cite{Glashow:1976nt}
\bibitem{Glashow:1976nt}
  S.L.~Glashow and S.~Weinberg,
  %``Natural Conservation Laws for Neutral Currents,''
  Phys.\ Rev.\ D {\bf 15} (1977) 1958.
  %%CITATION = PHRVA,D15,1958;%%
  %1296 citations counted in INSPIRE as of 17 Feb 2015

%\cite{Paschos:1976ay}
\bibitem{Paschos:1976ay}
  E.A.~Paschos,
  %``Diagonal Neutral Currents,''
  Phys.\ Rev.\ D {\bf 15} (1977) 1966.
  %%CITATION = PHRVA,D15,1966;%%
  %359 citations counted in INSPIRE as of 31 Mar 2015

\bibitem{Wiebusch} M. Wiebusch, ICHEP 2014. %%  these proceedings.

\bibitem{Kikuchi} M. Kikuchi, ICHEP 2014. %%  these proceedings.

%\cite{Pich:2009sp}
\bibitem{Pich:2009sp}
  A.~Pich and P.~Tuz\'on,
  %``Yukawa Alignment in the Two-Higgs-Doublet Model,''
  Phys.\ Rev.\ D {\bf 80} (2009) 091702.
%  [arXiv:0908.1554 [hep-ph]].
  %%CITATION = ARXIV:0908.1554;%%
  %141 citations counted in INSPIRE as of 17 Feb 2015

%\cite{Jung:2010ik}
\bibitem{Jung:2010ik}
  M.~Jung, A.~Pich and P.~Tuz\'on,
  %``Charged-Higgs phenomenology in the Aligned two-Higgs-doublet model,''
  JHEP {\bf 1011} (2010) 003.
 % [arXiv:1006.0470 [hep-ph]].
  %%CITATION = ARXIV:1006.0470;%%
  %81 citations counted in INSPIRE as of 17 Feb 2015

%\cite{Celis:2013ixa}
\bibitem{Celis:2013ixa}
  A.~Celis {\it et al.},  % V.~Ilisie and A.~Pich,
  %``Towards a general analysis of LHC data within two-Higgs-doublet models,''
  JHEP {\bf 1312} (2013) 095,
 % [arXiv:1310.7941 [hep-ph]].
  %%CITATION = ARXIV:1310.7941;%%
  %34 citations counted in INSPIRE as of 17 Feb 2015
%
%\cite{Celis:2013rcs}
%\bibitem{Celis:2013rcs}
%  A.~Celis, V.~Ilisie and A.~Pich,
  %``LHC constraints on two-Higgs doublet models,''
%  JHEP
{\bf 1307} (2013) 053.
 % [arXiv:1302.4022 [hep-ph]].
  %%CITATION = ARXIV:1302.4022;%%
  %74 citations counted in INSPIRE as of 14 Oct 2014

%\cite{CMS:2014xfa}
\bibitem{CMS:2014xfa} CMS and LHCb Collaborations,
  %``Observation of the rare $B^0_s\to\mu^+\mu^-$ decay from the combined analysis of CMS and LHCb data,''
  Nature {\bf 522} (2015) 68.
%  [arXiv:1411.4413 [hep-ex]].
  %%CITATION = ARXIV:1411.4413;%%
  %41 citations counted in INSPIRE as of 04 juin 2015

%\cite{Bobeth:2013uxa}
\bibitem{Bobeth:2013uxa}
  C.~Bobeth {\it et al.},
  %, M.~Gorbahn, T.~Hermann, M.~Misiak, E.~Stamou and M.~Steinhauser,
  %``B_{s,d} -> l+ l- in the Standard Model with Reduced Theoretical Uncertainty,''
  Phys.\ Rev.\ Lett.\  {\bf 112} (2014) 101801.
  % [arXiv:1311.0903 [hep-ph]].
  %%CITATION = ARXIV:1311.0903;%%
  %70 citations counted in INSPIRE as of 31 Mar 2015

%\cite{Li:2014fea}
\bibitem{Li:2014fea}
  X.Q.~Li, J.~Lu and A.~Pich,
  %``$B_{s,d}^0 \to \ell^+\ell^-$ Decays in the Aligned Two-Higgs-Doublet Model,''
  JHEP {\bf 1406} (2014) 022.
 % [arXiv:1404.5865 [hep-ph]].
  %%CITATION = ARXIV:1404.5865;%%
  %13 citations counted in INSPIRE as of 31 Mar 2015

%\cite{Aaij:2013qta}
\bibitem{Aaij:2013qta} LHCb Collaboration,
  %R.~Aaij {\it et al.}  [LHCb Collaboration],
  %``Measurement of Form-Factor-Independent Observables in the Decay $B^{0} \to K^{*0} \mu^+ \mu^-$,''
  Phys.\ Rev.\ Lett.\  {\bf 111} (2013) 191801;
  %[arXiv:1308.1707 [hep-ex]].
  %%CITATION = ARXIV:1308.1707;%%
  %119 citations counted in INSPIRE as of 31 Mar 2015
%
%\cite{LHCb:2015dla}
%\bibitem{LHCb:2015dla}
%  The LHCb Collaboration [LHCb Collaboration],
  %``Angular analysis of the $B^{0} \rightarrow K^{*0} \mu^{+} \mu^{-}$ decay,''
  LHCb-CONF-2015-002. %, CERN-LHCb-CONF-2015-002.
  %%CITATION = LHCB-CONF-2015-002, CERN-LHCB-CONF-2015-002;%%

%\cite{Descotes-Genon:2014uoa}
\bibitem{Descotes-Genon:2014uoa} %S.~Descotes-Genon {\it et al.},
  S.~Descotes-Genon, L.~Hofer, J.~Matias and J.~Virto,
  %``On the impact of power corrections in the prediction of $B \to K^*\mu^+\mu^-$ observables,''
  JHEP {\bf 1412} (2014) 125;
 % [arXiv:1407.8526 [hep-ph]].
  %%CITATION = ARXIV:1407.8526;%%
  %23 citations counted in INSPIRE as of 09 Apr 2015
%
%\cite{Descotes-Genon:2015xqa}
%\bibitem{Descotes-Genon:2015xqa}
%  S.~Descotes-Genon {\it et al.}, % L.~Hofer, J.~Matias and J.~Virto,
  %``Theoretical status of $B \to K^* \mu^+\mu^-$: The path towards New Physics,''
  arXiv:1503.03328 [hep-ph].
  %%CITATION = ARXIV:1503.03328;%%

%\cite{Altmannshofer:2014rta}
\bibitem{Altmannshofer:2014rta}
  W.~Altmannshofer and D.~M.~Straub,
  %``State of new physics in $b\to s$ transitions,''
  arXiv:1411.3161 [hep-ph],
  %%CITATION = ARXIV:1411.3161;%%
  %28 citations counted in INSPIRE as of 09 Apr 2015
%
%\cite{Altmannshofer:2015sma}
%\bibitem{Altmannshofer:2015sma}
%  W.~Altmannshofer and D.~M.~Straub,
  %``Implications of $b\to s$ measurements,''
  arXiv:1503.06199 [hep-ph].
  %%CITATION = ARXIV:1503.06199;%%
  %2 citations counted in INSPIRE as of 09 Apr 2015

%\cite{Aaij:2014ora}
\bibitem{Aaij:2014ora} LHCb Collaboration,
  %R.~Aaij {\it et al.}  [LHCb Collaboration],
  %``Test of lepton universality using $B^{+}\rightarrow K^{+}\ell^{+}\ell^{-}$ decays,''
  Phys.\ Rev.\ Lett.\  {\bf 113} (2014) 151601.
  %[arXiv:1406.6482 [hep-ex]].
  %%CITATION = ARXIV:1406.6482;%%
  %41 citations counted in INSPIRE as of 09 Apr 2015

%\cite{Bobeth:2011nj}
\bibitem{Bobeth:2011nj}
  C.~Bobeth {\it et al.}, %G.~Hiller, D.~van Dyk and C.~Wacker,
  %``The Decay B --> K l^+ l^- at Low Hadronic Recoil and Model-Independent Delta B = 1 Constraints,''
  JHEP {\bf 1201} (2012) 107.
 % [arXiv:1111.2558 [hep-ph]].
  %%CITATION = ARXIV:1111.2558;%%
  %92 citations counted in INSPIRE as of 09 Apr 2015

%\cite{Lees:2012xj}
\bibitem{Lees:2012xj} BaBar Collaboration,
  % J.~P.~Lees {\it et al.}  [BaBar Collaboration],
  %``Evidence for an excess of $\bar{B} \to D^{(*)} \tau^-\bar{\nu}_\tau$ decays,''
  Phys.\ Rev.\ Lett.\  {\bf 109} (2012) 101802.
  % [arXiv:1205.5442 [hep-ex]].
  %%CITATION = ARXIV:1205.5442;%%
  %197 citations counted in INSPIRE as of 09 Apr 2015

%\cite{Fajfer:2012jt}
\bibitem{Fajfer:2012jt}
  S.~Fajfer  {\it et al.}, % J.~F.~Kamenik, I.~Nisandzic and J.~Zupan,
  %``Implications of Lepton Flavor Universality Violations in B Decays,''
  Phys.\ Rev.\ Lett.\  {\bf 109} (2012) 161801;
 % [arXiv:1206.1872 [hep-ph]].
  %%CITATION = ARXIV:1206.1872;%%
  %67 citations counted in INSPIRE as of 09 Apr 2015
%
%\cite{Fajfer:2012vx}
%\bibitem{Fajfer:2012vx}
%  S.~Fajfer, J.~F.~Kamenik and I.~Nisandzic,
  %``On the $B \to D^* \tau \bar \nu_{\tau}$ Sensitivity to New Physics,''
  Phys.\ Rev.\ D {\bf 85} (2012) 094025.
 % [arXiv:1203.2654 [hep-ph]].
  %%CITATION = ARXIV:1203.2654;%%
  %100 citations counted in INSPIRE as of 09 Apr 2015

%\cite{Celis:2012dk}
\bibitem{Celis:2012dk}
  A.~Celis {\it et al.}, % M.~Jung, X.~Q.~Li and A.~Pich,
  %``Sensitivity to charged scalars in $\boldsymbol{B\to D^{(*)}\tau\nu_\tau}$ and $\boldsymbol{B\to\tau\nu_\tau}$ decays,''
  JHEP {\bf 1301} (2013) 054.
 % [arXiv:1210.8443 [hep-ph]].
  %%CITATION = ARXIV:1210.8443;%%
  %63 citations counted in INSPIRE as of 09 Apr 2015

%\cite{Abazov:2013uma}
\bibitem{Abazov:2013uma} D0 Collaboration,
  % V.~M.~Abazov {\it et al.}  [D0 Collaboration],
  %``Study of CP -violating charge asymmetries of single muons and like-sign dimuons in pp¯ collisions,''
  Phys.\ Rev.\ D {\bf 89} (2014) 1,  012002.
 % [arXiv:1310.0447 [hep-ex]].
  %%CITATION = ARXIV:1310.0447;%%
  %27 citations counted in INSPIRE as of 24 Apr 2015

\bibitem{Zito} M. Zito, ICHEP 2014. %%  these proceedings.

\bibitem{Goswami} S. Goswami, ICHEP 2014. %%  these proceedings.

\bibitem{Tortola} M. Tortola, ICHEP 2014. %%  these proceedings.

%\cite{An:2015rpe}
\bibitem{An:2015rpe} Daya Bay Collaboration,
  % F.~P.~An {\it et al.}  [Daya Bay Collaboration],
  %``A new measurement of antineutrino oscillation with the full detector configuration at Daya Bay,''
  arXiv:1505.03456 [hep-ex].
  %%CITATION = ARXIV:1505.03456;%%

\bibitem{Wang} W. Wang, ICHEP 2014. %%  these proceedings.

%\cite{Minkowski:1977sc}
\bibitem{Minkowski:1977sc}
  P.~Minkowski,
  %``$\mu \to e\gamma$ at a Rate of One Out of $10^{9}$ Muon Decays?,''
  Phys.\ Lett.\ B {\bf 67} (1977) 421.
  %%CITATION = PHLTA,B67,421;%%
  %2112 citations counted in INSPIRE as of 13 May 2015

\bibitem{GMRS:79} M.~Gell-Mann {\it et al.},
  %P. Ramond, CALT-68-709 (1979);
  % M.~Gell-Mann, P.~Ramond and R.~Slansky,
  % {\it Complex Spinors and Unified Theories},
  in {\it Supergravity}, eds. P. van Nieuwenhuizen and D.Z. Freedman
  (North Holland, Stony Brook 1979), p.~315

\bibitem{YA:79}
  T.~Yanagida, in  %Proc. {\it Workshop on
  {\it Unified Theory and Baryon Number in
    the Universe}, eds. O.~Sawada and A.~Sugamoto (KEK, 1979), p.~95

%\cite{Canetti:2012kh}
\bibitem{Canetti:2012kh}
  L.~Canetti {\it et al.}, % M.~Drewes, T.~Frossard and M.~Shaposhnikov,
  %``Dark Matter, Baryogenesis and Neutrino Oscillations from Right Handed Neutrinos,''
  Phys.\ Rev.\ D {\bf 87} (2013)  093006.
  % [arXiv:1208.4607 [hep-ph]].
  %%CITATION = ARXIV:1208.4607;%%
  %87 citations counted in INSPIRE as of 10 Apr 2015

%\cite{Weinberg:1979sa}
\bibitem{Weinberg:1979sa}
  S.~Weinberg,
  %``Baryon and Lepton Nonconserving Processes,''
  Phys.\ Rev.\ Lett.\  {\bf 43} (1979) 1566.
  %%CITATION = PRLTA,43,1566;%%
  %1030 citations counted in INSPIRE as of 10 Apr 2015

%\cite{Gonzalez-Garcia:2014bfa}
\bibitem{Gonzalez-Garcia:2014bfa}
  M.C.~Gonz\'alez-Garc\'{\i}a {\it et al.}, % M.~Maltoni and T.~Schwetz,
  %``Updated fit to three neutrino mixing: status of leptonic CP violation,''
  JHEP {\bf 1411} (2014) 052.
 % [arXiv:1409.5439 [hep-ph]].
  %%CITATION = ARXIV:1409.5439;%%
  %75 citations counted in INSPIRE as of 10 Apr 2015

%\cite{Khachatryan:2015kon}
\bibitem{Khachatryan:2015kon} CMS Collaboration,
 % V.~Khachatryan {\it et al.}  [CMS Collaboration],
  %``Search for lepton-flavour-violating decays of the Higgs boson,''
  arXiv:1502.07400 [hep-ex].
  %%CITATION = ARXIV:1502.07400;%%
  %8 citations counted in INSPIRE as of 11 Apr 2015

%\cite{Celis:2014roa}
\bibitem{Celis:2014roa}
  A.~Celis {\it et al.}, ICHEP 2014, %% these proceedings,       % V.~Cirigliano and E.~Passemar,
  %``Disentangling new physics contributions in lepton flavour violating tau decays,''
  arXiv:1409.4439 [hep-ph].
  %%CITATION = ARXIV:1409.4439;%%
  %5 citations counted in INSPIRE as of 09 Apr 2015

\bibitem{Wuerthwein} F. W\"urthwein, ICHEP 2014. %% these proceedings.

\bibitem{Kim} Y.-K. Kim, ICHEP 2014. %% these proceedings.

%\cite{Djouadi:2013lra}
\bibitem{Djouadi:2013lra}
  A.~Djouadi,
  %``Implications of the Higgs discovery for the MSSM,''
  Eur.\ Phys.\ J.\ C {\bf 74} (2014) 2704.
 % [arXiv:1311.0720 [hep-ph]].
  %%CITATION = ARXIV:1311.0720;%%
  %15 citations counted in INSPIRE as of 24 Apr 2015

%\cite{Hahn:2013ria}
\bibitem{Hahn:2013ria} T.~Hahn {\it et al.},
  % T.~Hahn, S.~Heinemeyer, W.~Hollik, H.~Rzehak and G.~Weiglein,
  %``High-Precision Predictions for the Light CP -Even Higgs Boson Mass of the Minimal Supersymmetric Standard Model,''
  Phys.\ Rev.\ Lett.\  {\bf 112} (2014) 14,  141801.
  % [arXiv:1312.4937 [hep-ph]].
  %%CITATION = ARXIV:1312.4937;%%
  %73 citations counted in INSPIRE as of 24 Apr 2015

%\cite{Bechtle:2014yna}
\bibitem{Bechtle:2014yna} P.~Bechtle {\it et al.}, ICHEP 2014,
  % P.~Bechtle, K.~Desch, H.~K.~Dreiner, M.~Hamer, M.~Krämer, B.~O'Leary, W.~Porod and B.~Sarrazin {\it et al.},
  %``How alive is constrained SUSY really?,''
  arXiv:1410.6035 [hep-ph].
  %%CITATION = ARXIV:1410.6035;%%
  %2 citations counted in INSPIRE as of 24 Apr 2015

\bibitem{Carlini} R. Carlini, ICHEP 2014. %% these proceedings.

%\cite{Buchmueller:2014yva}
\bibitem{Buchmueller:2014yva} O.~Buchmueller {\it et al.},
 % O.~Buchmueller, R.~Cavanaugh, M.~Citron, A.~De Roeck, M.~J.~Dolan, J.~R.~Ellis, H.~Flaecher and S.~Heinemeyer {\it et al.},
  %``The NUHM2 after LHC Run 1,''
  Eur.\ Phys.\ J.\ C {\bf 74} (2014) 12,  3212.
  % [arXiv:1408.4060 [hep-ph]].
  %%CITATION = ARXIV:1408.4060;%%
  %12 citations counted in INSPIRE as of 24 Apr 2015

%\cite{deVries:2015hva}
\bibitem{deVries:2015hva} K.J.~de Vries {\it et al.},
  %K.~J.~de Vries, E.~A.~Bagnaschi, O.~Buchmueller, R.~Cavanaugh, M.~Citron, A.~De Roeck, M.~J.~Dolan and J.~R.~Ellis {\it et al.},
  %``The pMSSM10 after LHC Run 1,''
  arXiv:1504.03260 [hep-ph].
  %%CITATION = ARXIV:1504.03260;%%
  %3 citations counted in INSPIRE as of 24 Apr 2015

%\cite{Cahill-Rowley:2014twa}
\bibitem{Cahill-Rowley:2014twa} M.~Cahill-Rowley {\it et al.},
  %M.~Cahill-Rowley, J.~L.~Hewett, A.~Ismail and T.~G.~Rizzo,
  %``Lessons and prospects from the pMSSM after LHC Run I,''
  Phys.\ Rev.\ D {\bf 91} (2015) 5,  055002.
  %[arXiv:1407.4130 [hep-ph]].
  %%CITATION = ARXIV:1407.4130;%%
  %7 citations counted in INSPIRE as of 24 Apr 2015

\bibitem{Carena} M. Carena, ICHEP 2014. %% these proceedings.

%\cite{Ade:2013zuv}
\bibitem{Ade:2013zuv} Planck Collaboration,
  % P.~A.~R.~Ade {\it et al.}  [Planck Collaboration],
  %``Planck 2013 results. XVI. Cosmological parameters,''
  Astron.\ Astrophys.\  {\bf 571} (2014) A16;
  % [arXiv:1303.5076 [astro-ph.CO]].
  %%CITATION = ARXIV:1303.5076;%%
  %3387 citations counted in INSPIRE as of 24 Apr 2015
%
%\cite{Ade:2015xua}
%\bibitem{Ade:2015xua} Planck Collaboration,
  %P.~A.~R.~Ade {\it et al.}  [Planck Collaboration],
  %``Planck 2015 results. XIII. Cosmological parameters,''
  arXiv:1502.01589 [astro-ph.CO].
  %%CITATION = ARXIV:1502.01589;%%
  %189 citations counted in INSPIRE as of 24 Apr 2015


\bibitem{Martinez} E. Mart\'{\i}nez-Gonz\'alez, ICHEP 2014. %% these proceedings.

\bibitem{Berdugo} J. Berdugo, ICHEP 2014. %% these proceedings.

\bibitem{Stegmann} C. Stegmann, ICHEP 2014. %% these proceedings.

\bibitem{Araujo} H. Ara\'ujo, ICHEP 2014. %% these proceedings.

\bibitem{Sokolowska} D. Soko{\l}owska and B. \'Swie\.{z}ewska, ICHEP 2014. %% these proceedings.

\bibitem{Lindner} M. Lindner, ICHEP 2014. %% these proceedings.

\bibitem{Palladino} A. Palladino, ICHEP 2014. %% these proceedings.

\bibitem{Matthews} J.M. Matthews, ICHEP 2014. %% these proceedings.

\bibitem{Kappes} A. Kappes, ICHEP 2014. %% these proceedings.

%\cite{Lesgourgues:2014zoa}
\bibitem{Lesgourgues:2014zoa}
  J.~Lesgourgues and S.~Pastor,
  %``Neutrino cosmology and Planck,''
  New J.\ Phys.\  {\bf 16} (2014) 065002.
  % [arXiv:1404.1740 [hep-ph]].
  %%CITATION = ARXIV:1404.1740;%%
  %12 citations counted in INSPIRE as of 24 Apr 2015

\bibitem{Obrient} R. O'Brient, ICHEP 2014. %% these proceedings.

%\cite{Ade:2014xna}
\bibitem{Ade:2014xna} BICEP2 Collaboration,
 % P.~A.~R.~Ade {\it et al.}  [BICEP2 Collaboration],
  %``Detection of $B$-Mode Polarization at Degree Angular Scales by BICEP2,''
  Phys.\ Rev.\ Lett.\  {\bf 112} (2014) 24,  241101.
  %[arXiv:1403.3985 [astro-ph.CO]].
  %%CITATION = ARXIV:1403.3985;%%
  %1039 citations counted in INSPIRE as of 24 Apr 2015

%\cite{Ade:2015tva}
\bibitem{Ade:2015tva} BICEP2 and Planck Collaborations,
 % P.~A.~R.~Ade {\it et al.}  [BICEP2 and Planck Collaborations],
  %``Joint Analysis of BICEP2/$Keck ?Array$ and $Planck$ Data,''
  Phys.\ Rev.\ Lett.\  {\bf 114} (2015) 10,  101301.
  % [arXiv:1502.00612 [astro-ph.CO]].
  %%CITATION = ARXIV:1502.00612;%%
  %88 citations counted in INSPIRE as of 24 Apr 2015

\bibitem{Guth} A. Guth, ICHEP 2014. %% these proceedings.

%\bibitem{Barenboim} G. Barenboim, ICHEP 2014. %% these proceedings.
%\cite{Barenboim:2014tca}
\bibitem{Barenboim:2014tca}
  G.~Barenboim and O.~Vives, ICHEP 2014,
  %``Transplanckian masses in inflation,''
  arXiv:1405.6498 [hep-ph].
  %%CITATION = ARXIV:1405.6498;%%
  %1 citations counted in INSPIRE as of 24 Apr 2015

%\cite{Bezrukov:2014bra}
\bibitem{Bezrukov:2014bra}
  F.~Bezrukov and M.~Shaposhnikov,
  %``Higgs inflation at the critical point,''
  Phys.\ Lett.\ B {\bf 734} (2014) 249.
 % [arXiv:1403.6078 [hep-ph]].
  %%CITATION = ARXIV:1403.6078;%%
  %49 citations counted in INSPIRE as of 24 Apr 2015

%\cite{Hamada:2014iga}
\bibitem{Hamada:2014iga} Y.~Hamada {\it et al.},
  % Y.~Hamada, H.~Kawai, K.~y.~Oda and S.~C.~Park,
  %``Higgs Inflation is Still Alive after the Results from BICEP2,''
  Phys.\ Rev.\ Lett.\  {\bf 112} (2014) 24,  241301.
 % [arXiv:1403.5043 [hep-ph]].
  %%CITATION = ARXIV:1403.5043;%%
  %66 citations counted in INSPIRE as of 24 Apr 2015

%\cite{Pich:1998xt}
\bibitem{Pich:1998xt}
  A.~Pich,
  %``Effective field theory: Course,''
  hep-ph/9806303.
  %%CITATION = HEP-PH/9806303;%%
  %199 citations counted in INSPIRE as of 24 Apr 2015

%\cite{Grzadkowski:2010es}
\bibitem{Grzadkowski:2010es} B.~Grzadkowski {\it et al.},
  %B.~Grzadkowski, M.~Iskrzynski, M.~Misiak and J.~Rosiek,
  %``Dimension-Six Terms in the Standard Model Lagrangian,''
  JHEP {\bf 1010} (2010) 085.
  %[arXiv:1008.4884 [hep-ph]].
  %%CITATION = ARXIV:1008.4884;%%
  %286 citations counted in INSPIRE as of 24 Apr 2015

%\cite{Buchmuller:1985jz}
\bibitem{Buchmuller:1985jz}
  W.~Buchm\"uller and D.~Wyler,
  %``Effective Lagrangian Analysis of New Interactions and Flavor Conservation,''
  Nucl.\ Phys.\ B {\bf 268} (1986) 621.
  %%CITATION = NUPHA,B268,621;%%
  %980 citations counted in INSPIRE as of 24 Apr 2015

%\cite{Falkowski:2014tna}
\bibitem{Falkowski:2014tna}
  A.~Falkowski and F.~Riva,
  %``Model-independent precision constraints on dimension-6 operators,''
  JHEP {\bf 1502} (2015) 039.
  %[arXiv:1411.0669 [hep-ph]].
  %%CITATION = ARXIV:1411.0669;%%
  %15 citations counted in INSPIRE as of 24 Apr 2015

%\cite{Alonso:2013hga}
\bibitem{Alonso:2013hga} R.~Alonso {\it et al.},
  % R.~Alonso, E.~E.~Jenkins, A.~V.~Manohar and M.~Trott,
  %``Renormalization Group Evolution of the Standard Model Dimension Six Operators III: Gauge Coupling Dependence and Phenomenology,''
  JHEP {\bf 1404} (2014) 159.
 % [arXiv:1312.2014 [hep-ph]].
  %%CITATION = ARXIV:1312.2014;%%
  %51 citations counted in INSPIRE as of 24 Apr 2015

%\cite{Delgado:2013hxa}
\bibitem{Delgado:2013hxa} R.L.~Delgado {\it et al.},
  %R.~L.~Delgado, A.~Dobado and F.~J.~Llanes-Estrada,
  %``One-loop $W_LW_L$ and $Z_LZ_L$ scattering from the electroweak Chiral Lagrangian with a light Higgs-like scalar,''
  JHEP {\bf 1402} (2014) 121.
 % [arXiv:1311.5993 [hep-ph]].
  %%CITATION = ARXIV:1311.5993;%%
  %16 citations counted in INSPIRE as of 24 Apr 2015

%\cite{Espriu:2013fia}
\bibitem{Espriu:2013fia} D.~Espriu {\it et al.},
  %D.~Espriu, F.~Mescia and B.~Yencho,
  %``Radiative corrections to WL WL scattering in composite Higgs models,''
  Phys.\ Rev.\ D {\bf 88} (2013) 055002.
%  [arXiv:1307.2400 [hep-ph]].
  %%CITATION = ARXIV:1307.2400;%%
  %15 citations counted in INSPIRE as of 24 Apr 2015

%\cite{Aad:2014zda}
\bibitem{Aad:2014zda} ATLAS Collaboration,
  %G.~Aad {\it et al.}  [ATLAS Collaboration],
  %``Evidence for Electroweak Production of $W^{\pm}W^{\pm}jj$ in $pp$ Collisions at $\sqrt{s}=8$ TeV with the ATLAS Detector,''
  Phys.\ Rev.\ Lett.\  {\bf 113} (2014) 14,  141803.
  % [arXiv:1405.6241 [hep-ex]].
  %%CITATION = ARXIV:1405.6241;%%
  %22 citations counted in INSPIRE as of 24 Apr 2015


\end{thebibliography}

%% Authors are advised to use a BibTeX database file for their reference list.
%% The provided style file elsarticle-num.bst formats references in the required Procedia style

%% For references without a BibTeX database:

\end{document}